\documentclass{article}

\usepackage{arxiv}

\usepackage[utf8]{inputenc} 
\usepackage[T1]{fontenc}    
\usepackage[hidelinks]{hyperref}       
\usepackage{url}            
\usepackage{amsmath}
\usepackage{amssymb}
\usepackage{booktabs}       
\usepackage{amsfonts}       
\usepackage{nicefrac}       
\usepackage{microtype}      
\usepackage{lipsum}		
\usepackage{graphicx}
\usepackage{natbib}
\usepackage{doi}
\usepackage{acronym}                      
\usepackage{graphicx}
\usepackage{dcolumn}
\usepackage{bm}
\usepackage{xcolor}
\usepackage{booktabs}
\usepackage{bm}
\usepackage{subcaption}
\usepackage{siunitx}
\usepackage{acronym}                      
\usepackage{physics}
\newcommand{\Rey}{\mathrm{Re}}

\acrodef{PIV}[PIV]{Particle Image Velocimetry}
\acrodef{CFD}[CFD]{Computational Fluid Dynamics}
\acrodef{POD}[POD]{Proper Orthogonal Decomposition}
\acrodef{LOR}[LOR]{Low-Order Reconstruction}
\acrodef{ROM}[ROM]{Reduced-Order Model}
\acrodef{ROMs}[ROMs]{Reduced-Order Models}
\acrodef{SPOD}[SPOD]{Spectral Proper Orthogonal Decomposition}
\acrodef{DMD}[DMD]{Dynamic Mode Decomposition}
\acrodef{ISOMAP}[ISOMAP]{isometric mapping}
\acrodefplural{ROM}[ROMs]{Reduced-Order Models}
\acrodef{SVD}[SVD]{Singular Value Decomposition}
\acrodef{NTR}[NTR]{non-time-resolved}
\acrodef{PSD}[PSD]{Power Spectral Density}
\acrodef{VQPCA}{Vector Quantization Principal Component Analysis}
\acrodef{TLF}[TLF]{Two-Line Fit}
\acrodef{ST-CNM}[ST-CNM]{Space-Time Cluster-Based Network Model}
\acrodef{CNM}[CNM]{Cluster-based Network Model}
\acrodef{MDS}[MDS]{Multidimensional Scaling}

\title{\textbf{Divide and conquer complex flows.
\\Part I: cluster and manifold-based local analysis } }

\author{ \href{https://orcid.org/0009-0008-5601-9970}{\includegraphics[scale=0.06]{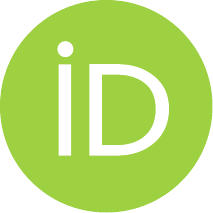}\hspace{1mm}Qihong L. Li-Hu}\\
    Department of Aerospace Engineering\\
    Universidad Carlos III de Madrid\\
     Leganés, Madrid, Spain\\
	\texttt{qihonglorena.li@uc3m.es} \\
	\And
	\href{https://orcid.org/0000-0001-7499-7569}{\includegraphics[scale=0.06]{orcid.pdf}\hspace{1mm}Guy Y. Cornejo Maceda} \\
	Department of Aerospace Engineering\\
    Universidad Carlos III de Madrid\\
     Leganés, Madrid, Spain\\
     \texttt{gcornejo@ing.uc3m.es}\\
    \And
	\href{https://orcid.org/0000-0001-7342-4814}{\includegraphics[scale=0.06]{orcid.pdf}\hspace{1mm}Andrea Ianiro} \\
	Department of Aerospace Engineering\\
    Universidad Carlos III de Madrid\\
     Leganés, Madrid, Spain\\
    \And
	\href{https://orcid.org/0000-0001-9025-1505}{\includegraphics[scale=0.06]{orcid.pdf}\hspace{1mm}Stefano Discetti} \\
	Department of Aerospace Engineering\\
    Universidad Carlos III de Madrid\\
     Leganés, Madrid, Spain\\
}

\renewcommand{\shorttitle}{\textit{arXiv} Template}

\hypersetup{
pdftitle={Divide and conquer complex flows. Part I: cluster and manifold-based local analysis},
pdfsubject={physics.flu-dyn},
pdfauthor={Qihong L. Li-Hu, Guy Y. Cornejo Maceda, Andrea Ianiro, Stefano Discetti},
pdfkeywords={Clustering, Manifold learning, ROM, POD, PIV},
}

\begin{document}
\maketitle

\begin{abstract}
This work is a two-part study on the description and prediction of complex fluid flows through the partitioning of the flow domain.
In this first part, we propose a framework for a global description of the dynamics of complex flows via clustered spatial representations of the flow, isolating and identifying local dynamics, retrieving different \acp{ST-CNM}.
The key enabler is the partitioning of the domain based on a nonlinear manifold learning approach, in which spatial points are clustered based on the similarity of their dynamics, as observed in their compact embedding in manifold coordinates.
The method receives as input time-resolved flow fields. The spatial manifold is computed through isometric mapping applied to the vorticity time histories at each spatial location.
An unsupervised clustering method, applied in the manifold space, partitions the full flow domain into subdomains.
The dynamics of each subdomain are then described with cluster-based modelling.
The method is demonstrated on two flow-field datasets obtained with a direct numerical simulation of a fluidic pinball under periodic forcing and with two-dimensional particle image velocimetry measurements of a transitional jet flow.
The spatial manifold-based flow partitioning identifies regions with similar dynamics in an automated way. For both cases, \ac{ST-CNM}  identifies local dynamics that are not captured by a global approach. In particular, vortex shedding and vortex pairing dynamics are isolated in the jet flow experiment. 
The proposed fully automated domain partitioning method will benefit the structural description of controlled flows and unveil the actuation mechanisms at play.
\end{abstract}


\section{Introduction} \label{sec:Introduction}

This work proposes a framework for a simple, local, and interpretable description of complex flows through atomised views of the dynamics. Despite the recent developments and increase in computational power, the high complexity of the models and the prohibitively expensive numerical simulations still comprise a large bottleneck when modelling turbulent and complex flows. Moreover, obtaining interpretable descriptions of the flow is critical for modelling, design, and control applications to optimise the performance and efficiency of many devices. In other words, an interpretable description of the flow consists of providing its structural organisation, i.e., the identification of different flow regions such as mixing layers or recirculation zones, of the main coherent features in each region, and of the regions of strong nonlinear interaction, including vortex tilting, stretching, and merging. Characterising these coherent structures and their intrinsic dynamics is essential for flow estimation and prediction. Understanding their origin and evolution is thus indispensable for flow design and control. In closed-loop control, the localisation of these structures and their evolution under forcing is fundamental for sensor and actuator placement, and for deriving actuation strategies targeting the dominant dynamical degrees of freedom. Moreover, in the era of machine-learning flow control~\citep{Duriez2017book,Rabault2019jfm,Brunton2020arfm}, where control laws or agents are often learned without explicit physical prior, an automated physical interpretation of the learned solutions is needed. For instance, the description of the evolution of coherent structures under actuation is essential to unveil the mechanisms discovered by the learning algorithm and to enable generalisation across different operating conditions and geometries.

The modelling of complex flows is challenged by their multiscale character, broadband frequency content, and intricate nonlinear interactions. Such features hinder the construction of \acp{ROM} that remain faithful, stable, and interpretable across regimes. With the increase in computational power and the development of advanced experimental diagnostics, data-driven modelling has emerged as a compelling alternative for extracting structure from high-dimensional flow fields. \ac{POD}~\citep{Lumley1967} has been widely used for this purpose, yet at high Reynolds numbers, the wide range of spatial and temporal scales necessitates a large number of modes, reducing interpretability. Projection-based reduced-order models built from such bases often become computationally expensive and exhibit complex nonlinear modal interactions that obscure physical insight~\citep{LiHu2025prf}. Methods designed to promote dynamical relevance, such as \ac{SPOD}~\citep{Towne2018jfm}, \ac{DMD}~\citep{Schmid2010jfm}, or multiscale \ac{POD}~\citep{Mendez2019jfm}, provide frequency- or time-scale-specific modes but rely on prior assumptions about the dominant time scales and do not naturally yield spatially localised structures. Autoencoders offer a flexible approach for nonlinear dimensionality reduction and prediction, but the learned latent variables and decoder modes usually remain difficult to interpret. Cluster-based modelling~\citep{Kaiser2014jfm,LiH2021jfm,Fernex2021sa,Deng2022jfm,HouC2022pof,HouC2024jfm} has proven effective in identifying quasi-attractors and transition processes by partitioning the trajectory in time. However, global clustering approaches are not naturally suited for resolving spatially localised coherent structures without introducing a prohibitively large number of clusters.


A promising approach to improve the fidelity and interoperability of models is to partition the data into coherent subsets, yielding local models. This idea is rooted in the philosophy of
\textit{divide et impera}
(divide and conquer) of
complex fluid flows, whereby 
a flow exhibiting coherent structures with distinct dynamics or under control
is decomposed into 
kinematically homogeneous subdomains
that can be modelled independently while retaining the global consistency. In this work, this philosophy is declined into two  flavours: kinematical interpretation (Part I) and dynamical prediction (Part II)~\citep{CornejoMaceda2026manuscript}.

The partitioning can be performed in the phase space and combined with intrusive model methods~\citep{colanera2025quantized}, in the sequence of snapshots combined with non-intrusive methods~\citep{Geelen2022ptrsa,Deng2025ams}, in the solution space~\citep{amsallem2012nonlinear} and in the spatial domain.
Spatially local approaches have recently been shown to improve generalisability and to reduce computational cost compared to full-domain models~\citep{Farcas2024aiaaj,Xu2025pof}. Domain partitioning has been used in \ac{CFD} to reduce computational cost, either through manual subdivision exploiting problem symmetries~\citep{Farcas2024aiaaj} or through criteria that define boundaries between regions of distinct dynamical behaviour. Examples include decompositions based on error indicators that distinguish low- and high-fidelity regions~\citep{Bergmann2018jcp}, and partitioning according to the decay of singular values to build locally coupled models~\citep{Gkimisis2025cmame}.
Finally, 
extended physics-informed neural networks (XPINNs)~\citep{Jagtap2020ccp}, the domain-partitioned version of PINNs, outperforms classical PINNs in terms of generalisation and convergence speed for the modelling of multiscale and multiphysics problems~\citep{HuZY2022siamjsc}.


Supervised and unsupervised machine learning methods have also been employed to decompose the domains.
Supervised methods aim to partition the domain based on a given flow characteristic.
For instance, \citet{LiBL2020jfm} trains a discriminator in a supervised way to separate turbulent and non-turbulent regions.
On the other hand, unsupervised flow partition aims to automatically extract relevant flow features.
These approaches rely on the definition of a metric
between spatial points that defines a feature space where the dynamics of the flow are separated.
Those clustering features have been defined in several ways. For example, \citet{Callaham2021nc} propose a clustering based on an equation space to identify local balances, thus approximating the equation of the dynamics in different regions of the space by a subset of the terms of the original equation.

Similarly, \citet{Otmani2023pof,Otmani2025ewc} separate viscous and turbulent regions using invariants of the strain-rate and rotation tensors.

\citet{parente2009identification} proposes another approach based on vector quantised principal component analysis (VQPCA) \citep{kambhatla1997dimension}, where the metric is defined as the reconstruction error of the local time series.
\citet{Munoz2023pof} compare different reduced order modelling methods (e.g., POD, autoencoders) and shows that VQPCA achieves the lowest reconstruction error.
Overall, the resulting domain partitioning obtained from these methods is expected to depend on the physics embedded in the features.

Moreover, methods based on causal analysis have been widely studied to identify and isolate the dominant mechanisms responsible for specific flow dynamics across different regions. This is achieved through SHAP analysis~\citep{cremades2024identifying} or quantifying information transfer and identifying the terms that contain the most relevant information. Works such as those of \cite{lozano2020causality, lozano2021cause,lozano2022information}, which are grounded in causal analysis, aim to interpret and develop reduced-order subgrid-scale models, thereby alleviating the computational cost associated with highly turbulent and multi-scale flows. Nonetheless, these methods rely on complete knowledge of all terms in the governing equations or on intrusive inference techniques that cannot be applied to experimental data.
Here, we propose a least-biased partitioning of the domain, i.e., based solely on snapshot data and with minimal human intervention. Moreover, modelling dynamical systems is essential for a comprehensive understanding of complex flows, as well as for applications such as flow control and performance optimisation. This constitutes the main objective of Part II~\citep{CornejoMaceda2026manuscript}, which focuses on the development of local dynamical models.

In the light of literature, the scientific gap we aim to fulfil is a least-biased interpretation of complex flows in a fully data-driven manner.
To this end,
we introduce an unsupervised domain partitioning method to model the flow based on a space-time clustering of snapshot data.
First, the domain is partitioned within a feature space (\emph{spatial manifold}) constructed from local time-series signatures of the flow field with \ac{ISOMAP}~\citep{Tenenbaum2000s}.
The novelty compared to previous studies lies in the definition of the features based on a nonlinear dimensionality reduction method.
Under the manifold hypothesis, high-dimensional flow data can be assumed to lie near a lower-dimensional structure parameterised by spatially varying local dynamics. Clustering within this spatial manifold yields an atomised representation of the dynamics, enabling the identification of spatially localised coherent structures and their local evolution.
Second, each subdomain is described with local Markov models based on cluster-based network modelling~\citep{LiH2021jfm,Fernex2021sa}.
The proposed method is demonstrated for two statistically stationary flows, a numerical simulation of the fluidic pinball under control~\citep{Deng2020jfm,CornejoMaceda2021jfm} and an experimental jet flow at Reynolds number  $\Rey=3300$~\citep{franceschelli2025assessment}. 

\subsection{Present study}
This study is split into two parts. Part I targets the identification of spatial subdomains with similar local kinematics. 
First, the spatial and temporal clustering methodology is presented in \S\ref{sec:methodology}.
The method is evaluated on a synthetic dataset of a controlled fluidic pinball, analysed in \S\ref{Sec:FluidicPinball}, and on the experimental measurements of a free jet flow in \S\ref{sec:JetFlow}.
Finally, \S\ref{Sec:Conclusion} summarises the results and offers concluding remarks.
The proposed domain partitioning methodology for fluid flows paves the way for compound local modelling, which is explored in \citet{CornejoMaceda2026manuscript}, the second part of this work.
The main hypothesis is that, once regions with similar physics are distilled and clustered, more compact models can accurately describe the flow behaviour.

\section{Methodology} \label{sec:methodology}
This section details the methodology of the \ac{ST-CNM}. First, an overview of the method is presented (\S\ref{Sec:Overview}). The domain decomposition and local cluster-based modelling are further explained in \S\ref{Sec:SpatialAnalysis} and \S\ref{Sec:CBModeling}.

\begin{figure}[htbp!]
    \centering
    \includegraphics[width=\textwidth]{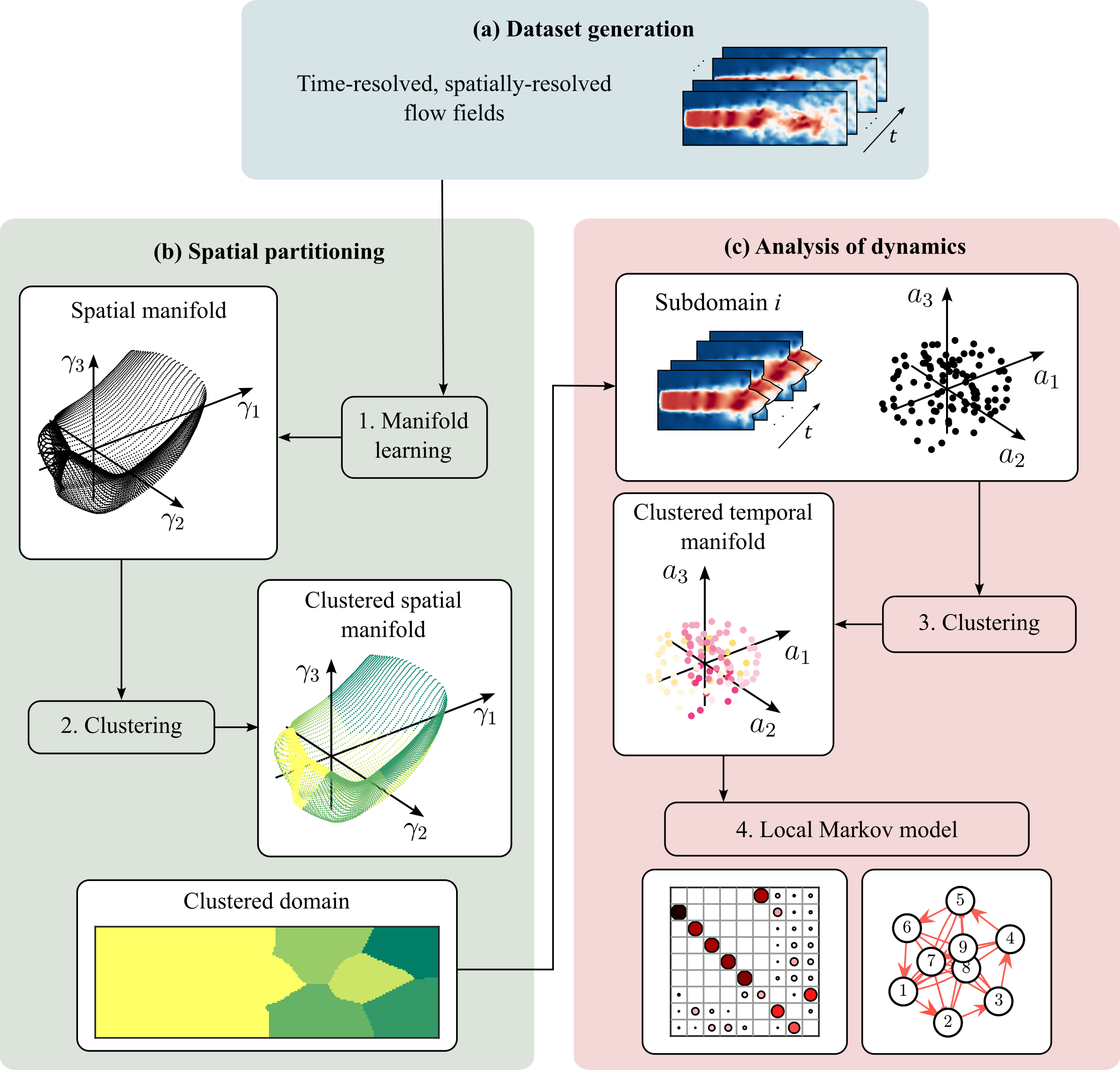}
    \caption{
Schematics of the space-time cluster-based network model methodology.
(a) Starting point is time-resolved velocity field snapshot data.
(b) The core idea being
building a \emph{spatial manifold}, i.e., deriving a set of coordinates where two points with similar dynamics are close to each other. $\gamma_i$ correspond to the spatial manifold coordinates.
The domain is then partitioned into subdomains with a clustering method applied to these spatial coordinates.
(c) Cluster-based kinematic models are built for each subdomain $i$, represented with cropped velocity field snapshots.
The atomised domain aids a simple interpretation of the dynamics.  $a_i$ correspond to the temporal modes of the flow fields.
}
    \label{Fig:Overview}
\end{figure}

\subsection{Overview}\label{Sec:Overview}
Figure~\ref{Fig:Overview} illustrates the three main steps of the methodology. 
\begin{enumerate}
    \item[(a)] The starting point is time-resolved and spatially-resolved snapshot data, typically obtained from numerical simulations or experimental measurements, such as \ac{PIV}.
\item[(b)] Then, the flow is partitioned in an unsupervised way into spatial subdomains.
\item[(c)] Finally, each subdomain is described with local Markov models with \acp{CNM}.
\end{enumerate}

The spatial partitioning step aims to separate regions based on local dynamics, so that regions with similar behaviour are assigned to the same subdomain.
The aim is to derive a dynamic-based partitioning
enabling the derivation of simple local models for each region.
The domain decomposition is carried out by first defining a feature space where two points with similar dynamics are close.
This space is built with an analysis of each grid point based on \ac{ISOMAP},
which is spanned by the coordinates $\gamma_1$, $\gamma_2$, and $\gamma_3$ in figure~\ref{Fig:Overview}.

The dimension of this space is determined by the minimum residual variance. 
Note that each point in this reduced space represents a point in the domain.
The shape described by all the points is referred to as \textit{spatial manifold}.
Then, a clustering algorithm is employed to partition the spatial manifold, and the affiliation function of each cluster is employed to partition the flow field into subdomains.
Note that the subdomains are not necessarily connected in the physical space, i.e, they represent cluster regions with similar dynamics even if physically disconnected. Further details are provided in Appendix \ref{App:JetSubdomains}.

To describe the dynamics of each subdomain, a temporal partitioning of the flow data is performed, where the snapshots correspond to the flow fields representing either the full domain or a subdomain.
For this, the cluster-based network modelling from \citet{LiH2021jfm} and \citet{Fernex2021sa} is employed.
First, the snapshots are partitioned into \textit{clusters} with a clustering algorithm.
The clusters correspond to the concept of clusters defined in \citet{LiH2021jfm}, and refer to a set of snapshots, without any specific ordering.
This term is chosen here to avoid any confusion with the clusters obtained in the spatial partitioning step, also referred to as a subdomain.
In the following, the term subdomain is employed exclusively for the spatial clusters.
On the other hand, cluster is used interchangeably for the classic temporal clusters.
Note that the clustering of the snapshots is equivalent to the clustering of the temporal \ac{POD} modes.
In practice, the \ac{POD} modes are preferred for their alleviated computational load~\citep{Kaiser2014jfm}.
Then, the transition probability matrix and transition time matrix are built based on the number of snapshots that leave from one cluster to another and the residence time in each cluster, respectively.
The first one describes a Markovian and stochastic transition of the flow from one cluster to another, while the latter gives the average transition time from one cluster to another.
To model the dynamics, each cluster is approximated by its centroid.
The validity of the model is tested by comparing the autocorrelation functions of the original flow and the predicted flow, obtained by sampling a new trajectory using the transition matrices from an initial centroid.
The time information is obtained from the transition time matrix, where the original time resolution is recovered by linear interpolation between successive centroids.

Before detailing the methodology of each step, we clarify the distinction between the space clustering performed in the first step, figure~\ref {Fig:Overview}(b), and the time clustering in the second step, figure~\ref{Fig:Overview}(c).
Clustering consists of partitioning a set of observations into clusters based on their similarity.
Without loss of generality, observations are described as vectors in an $n$-dimensional space, and the similarity between two vectors is measured by a prescribed metric, often based on the inner product of the space.
Finally, the clustering provides an affiliation function that, for each observation, retrieves its corresponding cluster.

For the spatial partitioning, one observation is the time sequence of a physical quantity at a given point in space.
When several quantities are considered, e.g., $u$, $v$, and $w$ components of a 3D flow, they are all concatenated to form one long sequence.
A cluster is a set of time sequence vectors, corresponding to several points in space.
The affiliation function is then employed to assign each space to a subdomain.
Contrary to the classic temporal clustering, the centroids of a space cluster do not have, a priori, a physical interpretation. Instead, they consist of the averaged time sequence of a physical quantity that includes phase shifts.
Note that temporal resolution is not required to compute the distance between time series; thus, this partitioning strategy can be applied to non-time-resolved data.

On the other hand, for the time partitioning of the data,
each observation is a snapshot at a given time.
A cluster is a set of snapshots at different times, and its centroid is the average flow field across all snapshots in the cluster.
The centroid corresponds to a representative realisation of the flow.
This clustering corresponds to the one introduced by \citet{Kaiser2014jfm}.
To avoid confusion between the time and space clustering,
the space clusters are referred to as subdomains,
while the time clusters are referred to as clusters.

\subsection{Dynamics-based spatial partitioning} \label{Sec:SpatialAnalysis}
This subsection describes the spatial partitioning of the flow.

%
Let us consider $N_t$ time- and space-resolved snapshots $\{\bm{u}(\bm{x},t_k) \}_{1 \leq k\leq N_t}$ of a velocity field.
We consider here a two-dimensional flow field for clarity, without loss of generality.
The velocity field is discretised on the grid defined by $N_s$ points $\bm{x}_i=(x_i,y_i) \in \mathbb{R}^{2}$.
The goal of the spatial partition is to derive non-trivial subsets of spatial points that cover the whole domain and whose intersection is empty.
A dynamics-based spatial partitioning aims to derive a spatial partitioning where points with similar dynamics are in the same subset.
The dynamics of each point are described by the time series of each physical quantity.
Following the clustering terminology, each point in space is an observation, and its associated time series is its features.

In this work, a vorticity time series is selected as the feature for each point as it synthesises both velocity components into a single quantity, thus alleviating the computational load.
The data matrix is then described as:
\begin{equation}
    \bm{W} = 
  \begin{bmatrix}
  \omega(\bm{x}_1,t_1) & \omega(\bm{x}_2,t_1) & \cdots & \omega(\bm{x}_{N_s},t_1) \\
  \omega(\bm{x}_1,t_2) & \omega(\bm{x}_2,t_2) & \cdots & \omega(\bm{x}_{N_s},t_2) \\
  \vdots          & \vdots          & \ddots & \vdots          \\
  \omega(\bm{x}_1,t_{N_t}) & \omega(\bm{x}_2,t_{N_t}) & \cdots & \omega(\bm{x}_{N_s},t_{N_t})
  \end{bmatrix}.
\end{equation}

Then, \ac{ISOMAP} is employed to define a space where two points are close to each other if their corresponding time series are close.
\ac{ISOMAP} consists of a nonlinear data decomposition method
that aims to extract a few coordinates that describe the dataset and conserve the geodesic distance between the points.
\ac{ISOMAP} has been recently employed in fluid mechanics to extract low-dimensional manifolds of the flow dynamics~\citep{farzamnik2023snapshots,Marra2024jfm}.
In this work, \ac{ISOMAP} is employed to construct a spatial manifold that maps the different types of dynamics in the flow.

The Euclidean distance $d_{i,j}$ between two time series $\omega_i = \omega(\bm{x}_i)$ and $\omega_j =\omega(\bm{x}_j)$ is computed according to:
\begin{equation}
    d_{i,j}
    = \int_0^T (\omega_i(t)-\omega_j(t))^2 \, dt.
\end{equation}
where $\bm{D} = [d_{i,j}]_{1\leq i,j \leq N_s}$ is the distance matrix.
In the following, a time series is also referred to as a data point.
Then, the manifold coordinates of the data are extracted with \ac{ISOMAP}.
The algorithm structure is briefly detailed here:
\begin{enumerate}
    \item \textbf{Build data network.}
    Each data point is linked to its $k$-nearest neighbours, forming one connected network $G$.
    
\item \textbf{Compute geodesic distance.} The geodesic distance between two points in $G$ is computed as the shortest path in the network. If two data points are neighbours, their distance corresponds to the Euclidean distance; otherwise, the shortest path is identified with the Floyd algorithm~\citep{Floyd1962algorithm}.
The resulting geodesic distance matrix is noted $\bm{D}_G \in \mathbb{R}^{N_s \times N_s}$.
\item \textbf{Compute \ac{ISOMAP} coordinates.} \ac{MDS}~\citep{Torgerson1952p}  is applied to $\bm{D}_G$ to extract a set of coordinates that span the low-dimensional embedding.
First, the matrix $\bm{D}_G$ is double-centred and diagonalised, with eigenvectors $\bm{w}_i$ and eigenvalues $\lambda_i$. 
The vector containing the $i$-th manifold coordinate of the points is defined as $\bm{\gamma}_i = \sqrt{\lambda_i}\, \bm{w}_i$.
In the following, the $i$-th dimension of the manifold is noted $\gamma_i$ (not bold).
\item \textbf{Select number of coordinates.} 
Following \citet{Tenenbaum2000s}, the number of dimensions retained is given by analysing the residual variance 
\begin{equation}
    R_v(n) = 1-R^2(\mathrm{vec}(\bm{D}_G),\mathrm{vec}(\bm{D}_n)),
    \label{Eq:ResidualVariance}
\end{equation}
where $R(\cdot,\cdot)$ is the correlation coefficient, $\mathrm{vec}(\cdot)$ is the vectorisation operator,
and $\bm{D}_n$ is the matrix of the Euclidean distances between the points in the manifold coordinate system, retaining only the first $n$ coordinates. This dimension $n$ is determined by an elbow criterion of the residual variance plot.
\end{enumerate}

We refer to \citet{Marra2024jfm} for more information on the \ac{ISOMAP} methodology.
The number of dimensions is chosen as the first minimum of $R_v(n)$, and the corresponding $n$-dimensional manifold extracted is referred to as the spatial manifold.
Each spatial point is represented by $n$ coordinates, which are the features employed for the clustering of the data.
The $k$-means algorithm employed for clustering takes as input the number of clusters $N_c$ and computes iteratively $N_c$ centroids that define a Voronoi tessellation of the $n$-dimensional space.
Each spatial point is represented by $n$ coordinates.
These coordinates are the features employed for clustering with methods such as $k$-means. 
$k$-means takes as input the number of clusters $N_c$ and computes iteratively $N_c$ centroids that define a Voronoi tessellation of the $n$-dimensional space.
Starting from the given $N_c$ initial centroids $\bm{c}_i$, the algorithm assigns each observation (in this case, each spatial point) to the closest centroid.
A set of data points, also known as a cluster, is then created for each centroid.
The cluster $i$ associated to centroid $\bm{c}_i$ is noted $\mathcal{C}_i$.
Then, the centroids are updated as the average within each cluster, and the algorithm is iterated until convergence.


The clustering method proposed in this work combines the $k$-means$++$ algorithm~\citep{Arthur2007} with the identification of fuzzy sets \citep{zadeh1965fuzzy}. The variant $k$-means$++$ optimises the initial centroids to maximise the distance between each other. It is a clustering technique where data belongs to unique, definite clusters.
However, to aid the global reconstruction of the dynamics,
the transition of coherent structures from one subdomain to another needs to be accounted for.
For this purpose, we propose subdomains with smooth transitions between each other through a method based on fuzzy sets
\citep{bezdek1973cluster,klir1994book,bezdek2013pattern}. Clustering based on fuzzy theory accounts for the measurement uncertainties while classifying. Thus, instead of having a hard partition of the data, it is classified into the so-called \textit{fuzzy sets}, which consist of sets or clusters with soft boundaries identified by grades of membership (in the interval [0,1]) to the given sets. This fuzzy methodology has been applied to clustering of image data for dimensionality reduction~\citep{sulaiman2010adaptive} and for partitioning algorithms~\citep{dunn1973fuzzy}, where adaptive boundaries and uncertainty-aware methods were required.

In fuzzy clustering, each point has a degree of belonging to the clusters, rather than belonging only to a unique cluster. Thus, points on the edge of a cluster may have a distribution of smaller degrees of belonging to the vicinity clusters than the points in the centre of the cluster. With this fuzzy $k$-means, the centroid of a cluster is computed as the mean of all points, weighting each point $x_i$ by its degree of belonging to the cluster $\mathcal{C}_k$, $W = w_{k} \in [0,1]$:
%
\begin{equation}
    \bm{c}_k =
    \cfrac{
        \iint_{\mathcal{D}} w_k (\bm{x})^m\; \bm{x} \; \textrm{d} \bm{x}
    }
    {
        \iint_{\mathcal{D}} w_k (\bm{x})^m \; \textrm{d} \bm{x}
    },
\end{equation}

\noindent where $k$ is the cluster index, and $m>1$ corresponds to the parameter controlling the level of fuzziness of the clustering. 

The degree of belonging is computed as:
%
\begin{equation}
    w_k (\bm{x}) = \frac{1}
    {
    \sum_{i=1}^{N_c}
    \left(
    \frac{d(\bm{c}_k,\bm{x})}
         {d(\bm{c}_i,\bm{x})}
    \right)^{2/(m-1)}
    }.
\end{equation}
which is normalised and fuzzified with the parameter $m$, and consists of a weighting factor of the distance of the points to the centroids. 

The clustering algorithm aims at minimising the weighted objective function:
%
\begin{equation}
    \mathcal{J}(W,C) =
    \sum_{k=1}^{N_c}
    \iint_{\mathcal{D}}
        w_{k}(\bm{x})^m\;\|\bm{x} - \bm{c}_k\|^2 \; \textrm{d} \bm{x}.
\end{equation}

\noindent where $\|\bm{x} - \mathbf{c}_j\|^2$ is the chosen distance measured between a data point $x_i$ and the centre of cluster $\mathcal{C}_j$.

The clustering is initialised with a $k$-means$++$ stage, where a first estimate of the centroids is computed. Then, the fuzzy clustering method is applied using as initial centroids the output from the $k$-means$++$.
The fuzzy clustering is achieved with the MATLAB function \texttt{fcm}.
Once the weights and centroids are derived, 
each point is assigned to the clusters based on a threshold. 
If a point has a weight for a given cluster above the threshold,
it is fully assigned to this cluster,
setting the weight to one and zeroing the weights for the other clusters.
Otherwise, the weights are summed (adding the largest one first) until the threshold is reached
and the point belongs to several clusters.
The selected weights are then normalised, and the remaining ones are set to zero. 

Note that the fuzziness and overlapping of clusters are useful in the identification and modelling of global dynamics, accounting for the transitional dynamics between domains. Thus, accounting for soft boundaries allows a smoother description of the dynamics at the cluster boundaries. 
Fuzzy $k$-means gives a more realistic partition of the flow field with regions where competing dynamics coexist.
This study focuses on the description of the local kinematics, thus well-defined clusters are sufficient.
In the second part of this work \citep{CornejoMaceda2026manuscript}, we exemplify how overlapping subdomains aid on the reconstruction of the global dynamics. 

The number of clusters is decided according to the \ac{TLF} method~\citep{Brindise2017eif} and they are ordered from larger to lower energy following the \ac{POD} convention.
This method consists of sweeping a large range of values and identifying the value that best splits the evolution of the explained variance into two linear segments.
First, the data is clustered for a large range of values, and the explained variance is computed.
The explained variance for $N_c$ clusters is defined as:
\begin{equation}
    E_v(N_c) = 1-\cfrac{\mathrm{WCSS}(N_c)}{\mathrm{WCSS}(1)}
\end{equation}
where $\mathrm{WCSS(N_c)}$ is the within-cluster sum of squares for $N_c$ clusters: 
$$\mathrm{WCSS}(N_c)=\sum\limits_{k=1}^{N_c} \cfrac{1}{|\mathcal{C}_k|} \sum\limits_{\bm{z} \in \mathcal{C}_k} \|\bm{z}-\bm{c}_k\|^2.$$
The number of clusters is typically computed until the convergence of the explained variance.
For the configurations considered in this work, the clustering is performed up to $50$ clusters.
Then, for a given number $N_c$ of clusters, the evolution of the explained variance is split in two parts, and the lower and upper parts are fit with linear functions.
The optimal number of clusters is chosen as the value that maximises the ratio $R^2/\epsilon$, where $R^2$ is the averaged coefficient of determination
and $\epsilon$ is the prediction error
defined as the sum of the difference between the predicted and the original values.
In the following, in the context of domain partitioning, the resulting clusters are referred to as subdomains and noted $\mathcal{D}_i$.

No meta-parameter tuning is required in the spatial partitioning presented in this section, with the exception of the number of neighbours needed in \ac{ISOMAP}.
Nonetheless, this parameter is expected to have a limited influence on the resulting manifold as long as the value is not so high that it causes a short-circuit in the neighbouring network.

\subsection{Local cluster-based network model} \label{Sec:CBModeling}
Once the domain is partitioned, a \ac{CNM}~\citep{LiH2021jfm} is built for each subdomain.
The methodology follows the steps of the original paper, i.e., (1) temporal clustering of the snapshots, (2) computation of the direct transition probability matrix, $\bm{Q}$, and the transition time matrix, $\bm{T}$.
Note that in this section, clustering corresponds to a partitioning of the snapshots, namely, extracting clusters of snapshots based on their similarities.
In this context, the observations are the snapshots and the features are the $u$ and $v$ components of the velocity at each point of the subdomain.
In short, the data is viewed as $N_t$ observations with $2\times N_{s,i}$ features each,
where $N_{s,i}$ is the number of points in subdomain $\mathcal{D}_i$.

Based on this description of the data, clustering is performed in the same way as in \S\ref{Sec:SpatialAnalysis}.
The metric between two snapshots $\bm{u}$ and $\bm{v}$ is then based on the inner product defined on the Hilbert space $\mathcal{L}^2(\mathcal{D}_i)$ of square-integrable vector fields in the domain $\mathcal{D}_i$,
\begin{equation}
    (\bm{u},\bm{v})_{\mathcal{D}_i} := \int_{\mathcal{D}_i} \bm{u} \cdot \bm{v} \, d\bm{x}.
\end{equation}
As mentioned in \S\ref{Sec:Overview}, the clustering can also be carried out on the temporal \ac{POD} modes.
In this case, the features of each observation are the temporal POD modes $a_i$.
In other words, for the $k$-th observation,
corresponding to snapshot $\bm{u}(\bm{x},t_k)$,
the features are $\bm{a}_k = (a_i(t_k))$,
Note that the temporal POD modes describe a ``temporal" manifold where the dynamics of the flow evolve.
The resulting temporal clusters are referred to as clusters and are noted $\mathcal{C}_j$.
The associated centroids are noted $\bm{c}_j$.
We recall that $\bm{c}_j$ is a snapshot as an average of the snapshots in $\mathcal{C}_j$.

Similarly to \S\ref{Sec:SpatialAnalysis}, the number of clusters is chosen with the \ac{TLF} criterion.

The direct transition probability $\bm{Q}$ and transition time matrices $\bm{T}$ are then computed.
The transition probability from $\mathcal{C}_j$ to $\mathcal{C}_i$ is:
\begin{equation}
    Q_{ij} = \cfrac{n_{ij}}{n_j}
\end{equation}
where $n_{ij}$ is the number of snapshots in $\mathcal{C}_j$ whose immediate successor in the original sequence is in $\mathcal{C}_i$,
and $n_j$ is the number of snapshots in $\mathcal{C}_j$ whose immediate successor in the original sequence is no longer in $\mathcal{C}_j$.
The transition time from $\mathcal{C}_j$ to $\mathcal{C}_i$ is 
\begin{equation}
    T_{ij} = <\cfrac{\tau_i+\tau_j}{2}>
\end{equation}
where $\tau_{j}$ is the residence time in $\mathcal{C}_j$, i.e., the number of successive snapshots times in $\mathcal{C}_j$ for a trajectory,
and $<\cdot>$ denotes the ensemble average over all trajectories.

The flow field data are then reconstructed by sampling a sequence of clusters from the transition probability matrix $\bm{c}_{k_0}$, $\bm{c}_{k_1}$, $\bm{c}_{k_2}$, etc., where now $k_i$ refers to the index of the temporal cluster.
The time information is retrieved with the transition time matrix: $t_0=0$, $t_1=T_{k_1k_0}$, $t_2=t_1+T_{k_2k_1}$, etc.
The reconstructed velocity field at $t \in [t_n, t_{n+1}]$ is then:
\begin{equation}
\begin{split}
    \hat{\bm{u}}(\bm{x},t)= \alpha_n \bm{c}_{k_n}(\bm{x}) + (1-\alpha_n(t)) \bm{c}_{k_{n+1}}(\bm{x})
    \\
    \alpha_n = \cfrac{t_{n+1}-t}{t_{n+1}-t_n}
\end{split}   
\end{equation}
Similar to \citet{LiH2021jfm}, the validity of the local cluster-based network models is verified with the autocorrelation function of the velocity field~\citep{Protas2015jfm}:
\begin{equation}
    R(\tau) = \cfrac{1}{T-\tau}
    \int_{\tau}^T
    \int_{\mathcal{D}_i}
    \bm{u}(\bm{x},t-\tau) \cdot \bm{u}(\bm{x},t)
    \, d\bm{x}\, dt, \quad \tau\in[0,T)
\end{equation}

The local \ac{CNM} graph from the transition probability matrix can be interpreted as a directed weighted Markov graph. Its nodes are representative local flow states, and its edges are empirical transition probabilities. Closed paths in the graph correspond to recurrent local dynamics, and their characteristic periods are obtained by summing the transition times along the cycle.  A graph in which each node has one dominant outgoing edge corresponds to a nearly deterministic periodic process, whereas multiple outgoing edges indicate intermittency or switching between different local dynamics. Nearly deterministic cycles indicate periodic local dynamics, whereas branching paths indicate switching between competing mechanisms. Since the present model uses nonuniform transition times, the physical periods are more directly inferred from the transition-time matrix and autocorrelation analysis.



\section{Flow description of the fluidic pinball under control}\label{Sec:FluidicPinball}
The \ac{ST-CNM} methodology is first demonstrated for a two-dimensional numerical dataset of a flow under control action, the fluidic pinball.
The flow is forced at different incommensurable frequencies to mimic a multi-frequency flow.

\subsection{Configuration}
\begin{figure}[htbp!]
    \centering
    \begin{subfigure}[t]{0.5\textwidth}
        \centering
        \includegraphics[width=\linewidth]{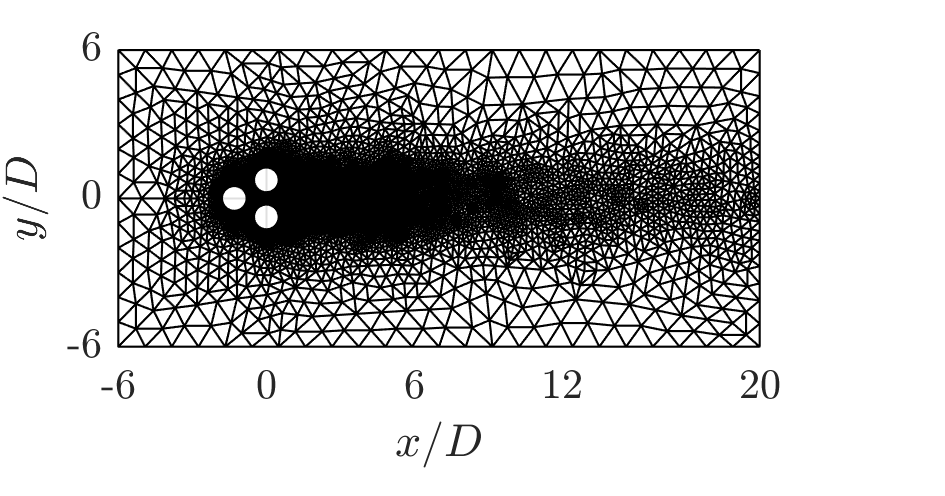}
        \caption{}
        \label{Fig:Grid}
    \end{subfigure}\hfill
    \begin{subfigure}[t]{0.5\textwidth}
        \centering
        \includegraphics[width=\linewidth]{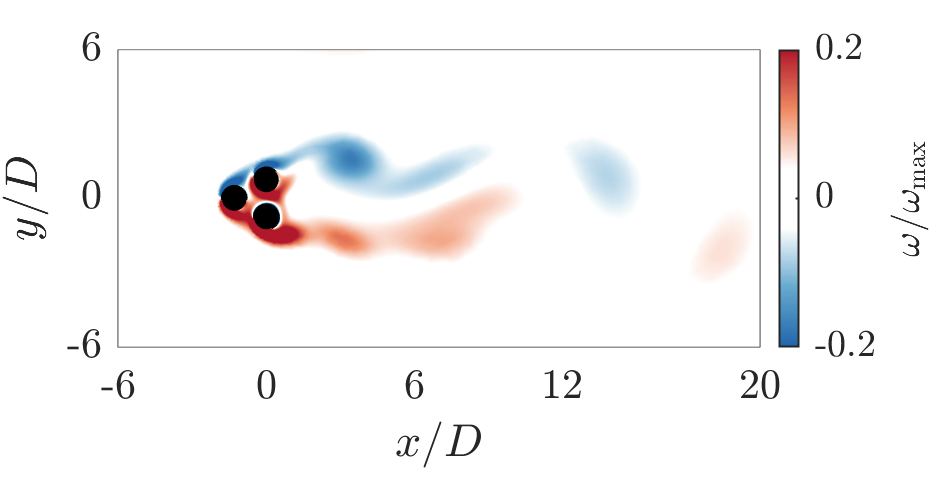}
        \caption{}
    \label{Fig:FP_snapshot}
    \end{subfigure}
    \caption{Fluidic pinball configuration. 
    (a) Computational grid. 
    (b) Vorticity field of a sample snapshot of the flow under control.}
    \label{Fig:FP_configuration}
\end{figure}

Figure~\ref{Fig:FP_configuration} depicts the fluidic pinball configuration as described in the work by \citet{Deng2020jfm}.
The system consists of three cylinders located at the vertices of an equilateral triangle pointing upstream.
The diameter of the cylinders is noted as $D$.
The centres of the cylinders are located at $3D/2$ of each other.
The flow is described in a Cartesian coordinate system.
The origin is located halfway between the two rear cylinders.
The $x$-axis is oriented with the direction of the incoming flow.
The $y$-axis is chosen such that the angle from the $x$- to the $y$-axis points counter-clockwise.
The flow domain extends 20$D$ downstream, 6$D$ upstream, and 12$D$ spanwise.

We consider an incompressible flow at $\Rey =U_{\infty}D / \nu=30$,
where $U_{\infty}$ is the incoming velocity,
and $\nu$ the kinematic viscosity of the fluid.
The density of the fluid is noted as $\rho$. 
All quantities are normalised by the $D$, $U_{\infty}$ and $\rho$.
At this Reynolds number, the flow is beyond the first Hopf bifurcation, and the only attractor in the phase space is a symmetric limit cycle~\citep{Deng2020jfm}.
Periodic vortex shedding forming a von K\'arm\'an vortex street is observed.
The non-dimensional frequency of the unforced vortex shedding is $f_0=0.088$ and the corresponding period $T_0=11.36$, normalised with the convective time.

The flow is studied through direct numerical simulations of the incompressible Navier-Stokes equations.
The numerical simulations are carried out with an in-house solver validated in the works by \citet{Deng2020jfm} and \citet{Noack2003jfm}.
The Navier-Stokes equations are solved based on a second-order finite element discretisation method on an unstructured grid of $4225$ triangles and $8633$ vertices, see figure~\ref{Fig:Grid}.
The method is third-order accurate in time and space.
The velocity is set to $\bm{u} = (U_{\infty},0)$ for the left, top, and bottom boundaries.
A stress-free condition is set for the right boundary.
The boundary conditions on the cylinders are updated each time step to simulate their rotation.

To mimic a multifrequency flow, the two rear cylinders are controlled with a periodic signal but with different frequencies.
The peripheral speed of the top and bottom cylinders is set to $b_{T} = \cos(2\pi f_{c1}t )$ and $b_{B} = \cos(2\pi f_{c2}t )$, respectively. The frequencies are set to
$f_{c1} = \sqrt{\pi} f_0$ 
and $f_{c2} = \pi f_0$,
and the corresponding periods are
$T_{c1} \approx 6.41$ and $T_{c2} \approx 3.62$. 
The $\pi$ and $\sqrt{\pi}$ coefficients are chosen so that the local dynamics of the flow are incommensurable with the natural vortex shedding.
For this study, $10000$ snapshots in the post-transient regime are sampled at $\Delta t = 0.1$.
This corresponds to more than $80$ periods of the unforced flow.
Figure~\ref{Fig:FP_snapshot} shows the vorticity field for one snapshot.

\definecolor{mygreen}{rgb}{0.25,0.75,0.25}
\definecolor{myblue}{rgb}{0.375,0.375,1}
\definecolor{myred}{rgb}{1,0.375,0.375}

\subsection{Spatial manifold}
\begin{figure}[htbp!]
    \centering
    \includegraphics[width = 0.75\textwidth]{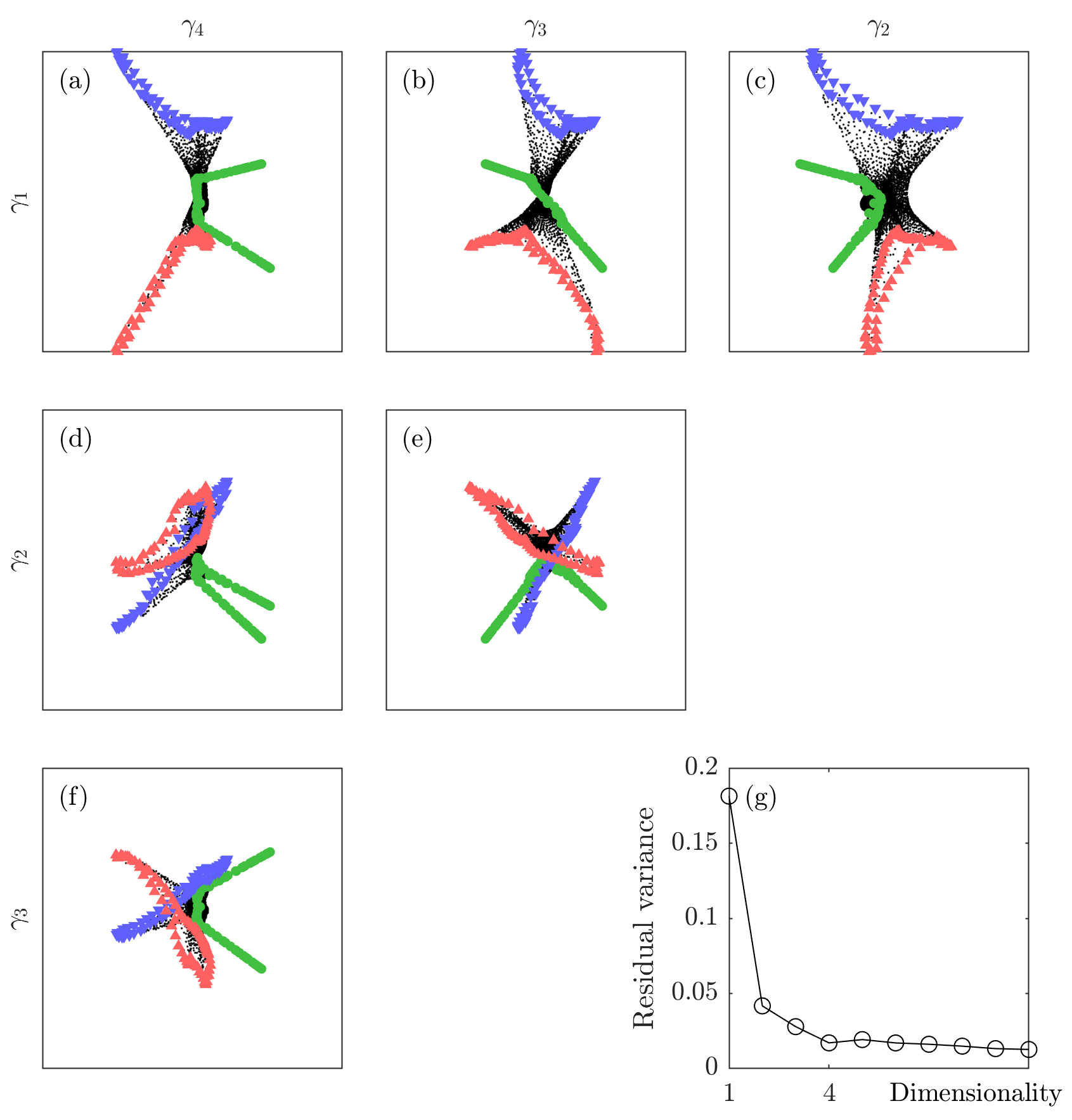}
    \caption{Spatial manifold for the fluidic pinball. (a-f) Projection of the manifold on the first four manifold coordinates.
    The points on the cylinders are indicated with ${\Huge \textcolor{mygreen}{\bullet}}$, $\textcolor{myblue}{\blacktriangledown}$, and $\textcolor{myred}{\blacktriangle}$ for the front, bottom, and top cylinders, respectively.
    The remaining points are indicated with black dots.
    (g) Residual variance as a function of the dimensionality of the manifold.
    }
    \label{Fig:FP_SpatialManifold}
\end{figure}
For spatial partitioning of the flow field, points on the unstructured grid are used directly.
This allows better separation of the subdomains in regions where the mesh is refined.
The spatial manifold is built following the methodology described in \S\ref{Sec:SpatialAnalysis}.
$k=10$ neighbours are considered during the \ac{ISOMAP} step to build the spatial manifold.
Figure~\ref{Fig:FP_SpatialManifold} describes the obtained manifold, where the residual variance presents one minimum for four subdomains.
Hence, only the first four manifold coordinates are considered.
The spatial manifold displays a butterfly-type shape, see figure~\ref{Fig:FP_SpatialManifold} (c).
Interestingly, the wings are carried on the external side by points on the rear cylinders.
The points on the front cylinder form the skeleton of both antennas as they connect in the centre of the manifold.

\begin{figure}[htbp!]
 \centering
    \includegraphics[width=0.8\textwidth]{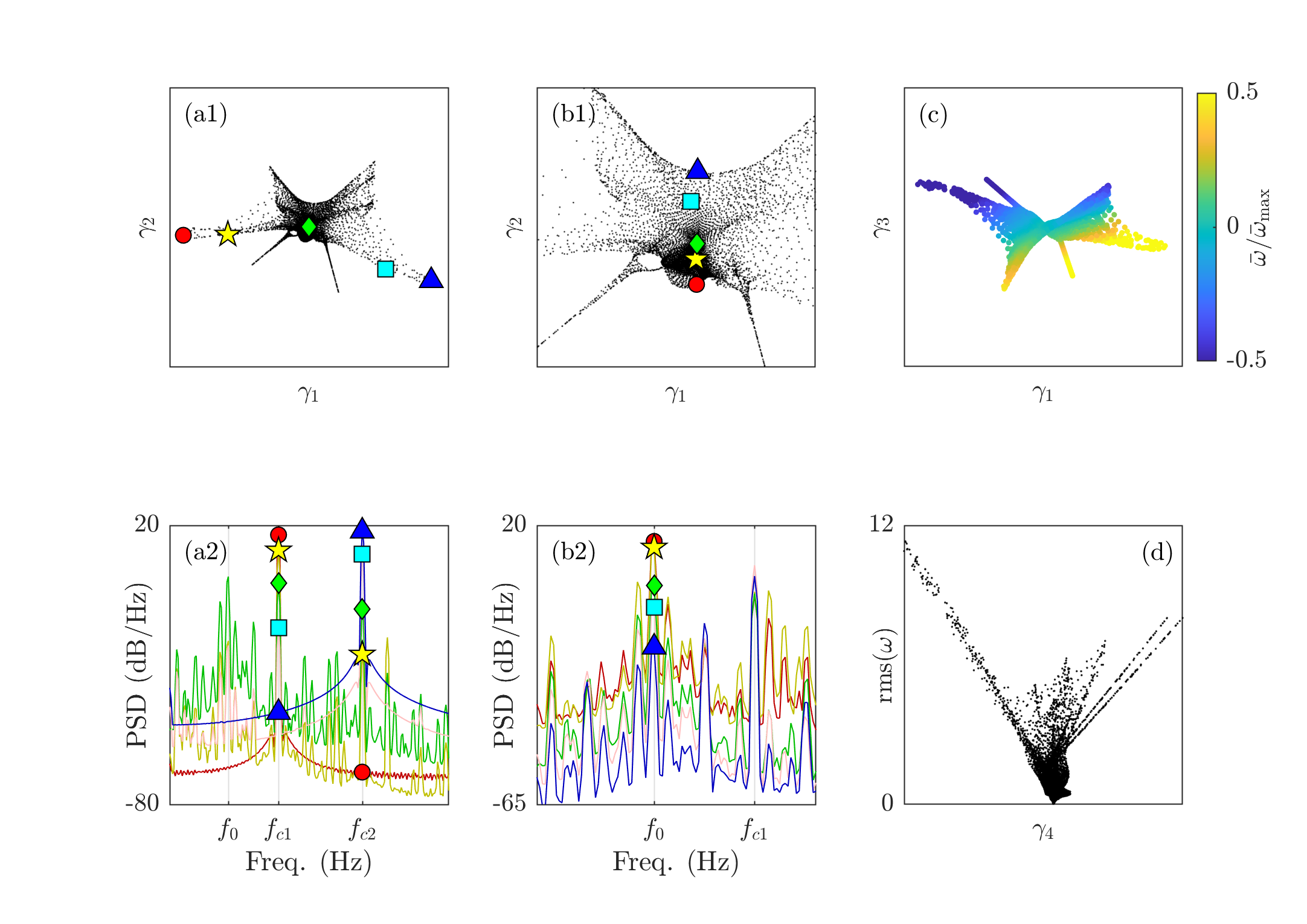}
    \caption{Intepretation of the spatial manifold coordinates $\gamma_i$ for the fluidic pinball. (a1) Selected points on the spatial manifold ``evenly'' sampled from maximum to minimum $\gamma_1$.
    (a2) PSD of the $v$-velocity of the points sampled in (a1). The symbols indicate the corresponding power levels for $f_{c1}$ and $f_{c2}$.
    (b) Similar to (a) for points sampled along $\gamma_2$.
    (c) Data points coloured by the mean vorticity.
    (d) Data points projected on the plane $\gamma_4-\mathrm{rms}(\omega)$.
    }
    \label{Fig:InterpretationGammas}
\end{figure}
Figure~\ref{Fig:InterpretationGammas} gives an interpretation of the manifold coordinates.
For the interpretation of $\gamma_1$, five evenly-spaced points along $\gamma_1$ are sampled.
First, the points with minimum and maximum $\gamma_1$ are selected.
The remaining points are the manifold points that are the closest to a linear interpolation between the two extremal ones.
The \ac{PSD} of each point is displayed in figure~\ref{Fig:InterpretationGammas}(a).
The results suggest that $\gamma_1$ is related to the spectral content of the local time series.
More precisely, $\gamma_1$ describes a continuous transition from $f_{c1}$ (low $\gamma_1$ values) to $f_{c2}$ (high $\gamma_1$ values).
A similar analysis carried out on the second manifold coordinate suggests that $\gamma_2$ is related to the power level of the natural frequency $f_0$.
For the third manifold coordinate, the scalar plot, figure~\ref{Fig:InterpretationGammas}(c), suggested that $\gamma_3$ is related to the mean vorticity of the points.
Finally, figure~\ref{Fig:InterpretationGammas} shows a piece-wise linear relationship between the root mean square value of the vorticity and $\gamma_4$.
The fourth manifold coordinate seems to describe a non-standard quantity whose interpretation needs further investigation.

\subsection{Local cluster-based models}
\begin{figure}[htbp!]
    \centering
    \includegraphics[width = \textwidth]{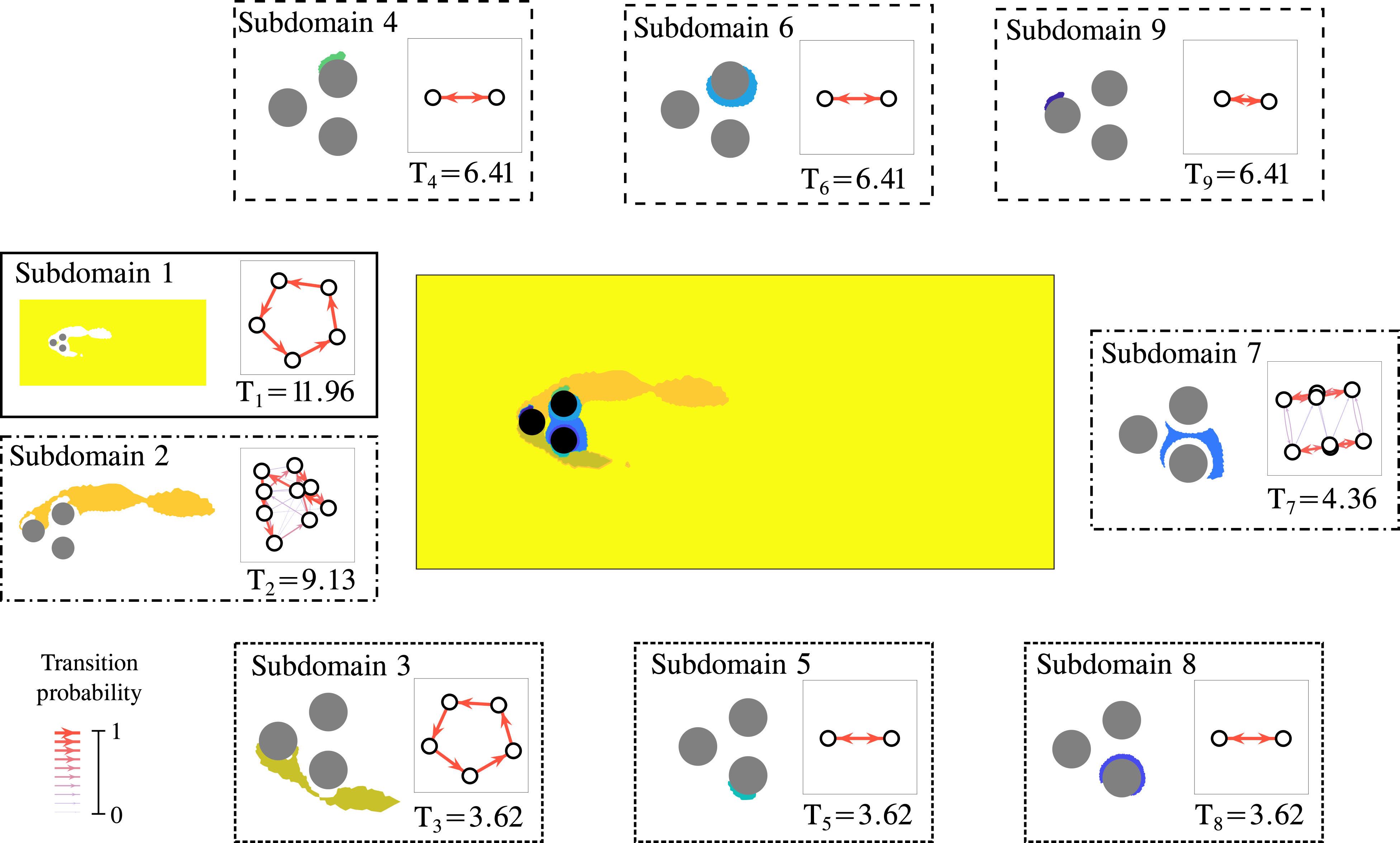}
    \caption{Local interpretation of the fluidic pinball under control.
    The flow is partitioned into $9$ subdomains.
    Each subdomain is accompanied by the cluster-based network model built on the masked snapshots.
    The nodes correspond to the centroids of the clusters projected on the first two POD modes. 
    The time $T_i$ corresponds to the main period of the flow obtained by computing the first zero of the autocorrelation function for the predicted data.
}
    \label{Fig:FP_Results}
\end{figure}
Figure~\ref{Fig:FP_Results} gives an exploded view of the fluidic pinball and the nine subdomains. The spatial partitioning is performed according to a criterion of regions exhibiting similar flow dynamics. First of all, three main subdomains can be identified. Subdomain $\mathcal{D}_1$ corresponds to the region where the flow is primarily dominated by advection, as flow enters the domain at a constant inflow velocity of $U_{\infty}$. This region extends throughout a large portion of the domain, including the downstream region of the fluidic pinball, where the dynamics remain largely convection-dominated despite the mixing induced by vortex shedding.

Additionally, two subdomains, $\mathcal{D}_2$ and $\mathcal{D}_3$, are identified along the shear layer generated by the front cylinder. Each side of the wake is grouped into a single subdomain due to the alternating dynamics associated with vortex shedding. The asymmetry between these regions appears to be influenced by the periodic actuation imposed on the rear cylinders, similarly to the forcing strategies investigated by \cite{he2000active}. The primary effect of the periodic forcing is observed in the near wake immediately downstream of the cylinders, whose extent depends on both the forcing frequency and amplitude. This behaviour is also reflected in the identified subdomains: subdomain $\mathcal{D}_2$ exhibits a longer shear-layer development, which is associated with a lower actuation frequency and therefore allows the shear layer originating from the front cylinder to persist further downstream. In contrast, subdomain $\mathcal{D}_3$ displays a shorter shear-layer region. In this case, the higher rotation frequency of the bottom cylinder appears to reduce the streamwise development of the shear layer. This behaviour may indicate that the influence of the actuation of the bottom cylinder is smaller than that of the top cylinder. Since both actuations are done with the same amplitude, varying the frequency, at higher frequencies, a higher amplitude would be required to obtain an effective control of the flow \citep{he2000active}.

Furthermore, several smaller subdomains are identified in the vicinity of the cylinders. Subdomains $\mathcal{D}_4$ and $\mathcal{D}_6$, as well as $\mathcal{D}_5$ and $\mathcal{D}_8$, exhibit similar dynamical behaviour, with dominant frequencies associated with the rotation of the corresponding cylinders. Similar to the observations reported by \citet{he2000active}, regions surrounding the cylinder surfaces can be identified, which are associated with flow separation and near-wake dynamics induced by vortex shedding. Nevertheless, it is important to note that the flow regime considered in the present work differs from that studied in \citep{he2000active}, although qualitatively similar features are observed.

In the present case, these regions are further partitioned into two parts, a smaller outer part ($\mathcal{D}_4$ and $\mathcal{D}_5$) and a longer inner region ($\mathcal{D}_6$ and $\mathcal{D}_8$). This subdivision may result from the interaction between the rear-cylinder actuation and the shedding generated by the front cylinder. The front cylinder also exhibits a small attached region (subdomain $\mathcal{D}_9$). Unlike the rear cylinders, the front cylinder is not actuated. However, this region appears to be influenced by the actuation of the top cylinder, whose rotation modifies the local flow velocity both upstream and downstream of the cylinder to a greater extent compared to the bottom cylinder, which has a more attenuated effect. Consequently, the dynamics in this region are strongly affected by the motion induced by the top cylinder.

Finally, subdomain $\mathcal{D}_7$ corresponds to a transition region characterised by strong interaction between the actuation of the top and bottom cylinders. Interestingly, this region appears to limit the development of coherent vortex shedding in the gap between the rear cylinders, leading to a significantly altered wake structure in the inner region.

The number of temporal clusters for the cluster-based network models is obtained via the \ac{TLF} method detailed in \S\ref{Sec:SpatialAnalysis}.
The method gives
five clusters for subdomains $\mathcal{D}_1$ and $\mathcal{D}_3$,
nine for subdomain $\mathcal{D}_2$,
two clusters for subdomains $\mathcal{D}_4$, $\mathcal{D}_5$, $\mathcal{D}_6$, $\mathcal{D}_8$, and $\mathcal{D}_9$,
and eight for subdomain $\mathcal{D}_7$.
An analysis of the autocorrelation of the original data and the prediction data shows an excellent overlap 
for all cases except subdomains $2$ and $7$ (not shown here).
The dominant period $T_i$ of the coherent structures is obtained through the first zero of the autocorrelation function. 
The periods are summarised in Table~\ref{Tab:FP_Periods}.

\begin{table}
    \centering
   \begin{tabular}{lccccccccc}
    \toprule
        Cluster \# & 1 & 2 & 3 & 4 & 5 & 6 & 7 & 8 & 9\\
        \midrule
        Original data & 11.65 & 8.64 & 3.79 & 6.42 & 3.62 & 6.39 & 4.19 & 3.64 & 6.34 \\
        \midrule
        CNM prediction & 11.96 & 9.13 & 3.62 & 6.41 & 3.62 & 6.41 & 4.36 & 3.62 & 6.41 \\
        \bottomrule
    \end{tabular}
    \caption{Dominant period for each subdomain of the fluidic pinball obtained via the first zero of the autocorrelation function.}
    \label{Tab:FP_Periods}
\end{table}

    
Three groups of subdomains stand out.
The first group includes subdomains $\mathcal{D}_4$, $\mathcal{D}_6$, and $\mathcal{D}_9$ (dashed boxes).
The period associated with these subdomains corresponds to the forcing period of the top cylinder ($T_{c1} = 6.41$).
Although these subdomains share similar dynamics, the spatial clustering separates them into distinct clusters.
This is explained by the average vorticity of these subdomains.
Indeed, we have seen that the third spatial feature $\gamma_3$ is correlated with the average vorticity, and the values for subdomains $\mathcal{D}_4$, $\mathcal{D}_6$, and $\mathcal{D}_9$ are $-5.31$, $1.48$, and $-6.28$, respectively.
Subdomains $\mathcal{D}_3$, $\mathcal{D}_5$, and $\mathcal{D}_8$ form the second group (dotted-line boxes).
The periods computed from the autocorrelation function correspond to the forcing period of the bottom cylinder, $T_{c2}$.
Similar to the first group, the distinction between the clusters is explained by the average vorticity level.
The last group (solid-line box) is the subdomain $\mathcal{D}_1$ alone.
This subdomain describes the far-field dynamics, whose dominant period is slightly greater than that of the unforced flow.
Subdomains $\mathcal{D}_2$ and $\mathcal{D}_7$ (dot-dashed-line box) display complex networks whose dominant period is not related to $f_0$, $f_{c1}$, or $f_{c2}$.
They are further interpreted below.

\begin{figure*}[t!]
    \centering
            \includegraphics[width = \textwidth]{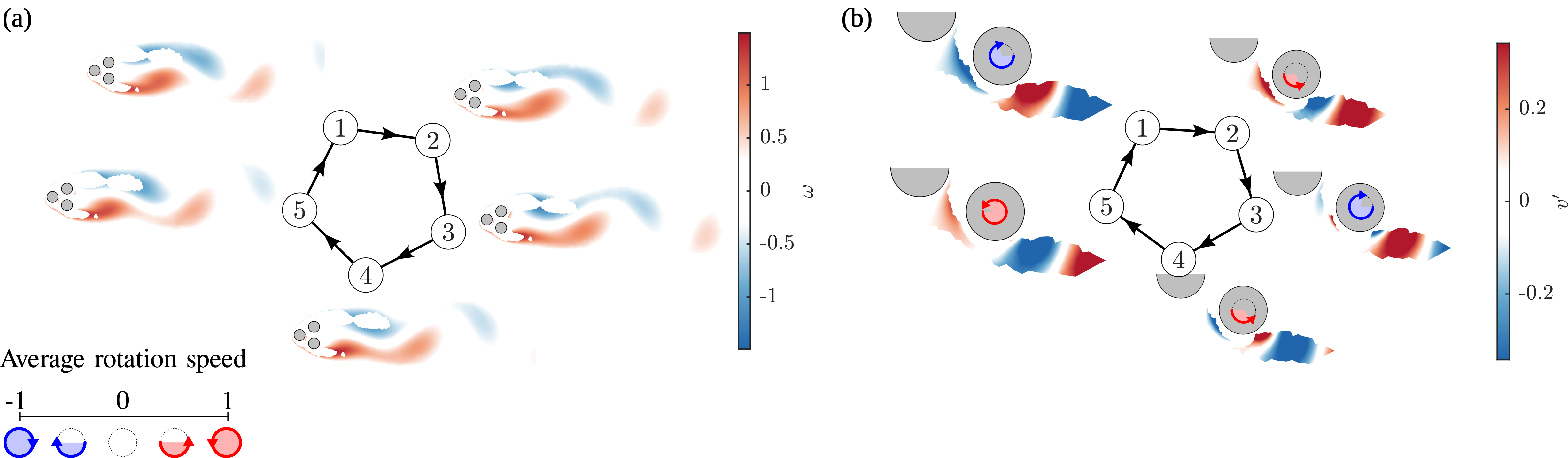}
    \caption{Description of $5$-clusters subdomains.
    Depiction of the centroids, transition and snapshots for (a) subdomain $\mathcal{D}_1$ and (b) subdomain $\mathcal{D}_3$.
    The circular arrows indicate the average rotation speed of the cylinder for each cluster.
    A full circle corresponds to a peripheral speed of 1 $D$ per convective time unit.
    }
    \label{Fig:FP_S7+S9}
\end{figure*}
Figure~\ref{Fig:FP_S7+S9} shows the centroids for the $5$-clusters subdomains.
Subdomain $\mathcal{D}_1$ represents the far-field where the dynamics are driven by the global vortex shedding.
The centroids are averaged vorticity fields in a cluster.
The sequence of snapshots describes the shedding of a vortex and its advection downstream.
Note that the averaged flow does not resolve the small structures in the vorticity branches.

On the other hand, for subdomain $\mathcal{D}_3$, the velocity fields show that the clusters are related to the rotation of the bottom cylinder.
The velocity changes sign between windward and leeward of the bottom cylinder, but keeps similar intensity.
Note that the velocity field indicates a displacement of the fluid in the opposite direction to the rotation of the cylinder.
For example, for cluster $\mathcal{C}_1$, the bottom cylinder rotates clockwise on average,
but the fluctuation of velocity is negative on the windward side and positive on the leeward side of the cylinder.
This is due to the phase delay between the generation of vorticity by the cylinder rotation and its propagation through viscous diffusion.
This description fits clusters $\mathcal{C}_1$, $\mathcal{C}_2$, and $\mathcal{C}_5$.
For clusters $\mathcal{C}_3$ and $\mathcal{C}_4$, the velocity field corresponds to the rotation direction.
Those clusters may correspond to phases of the flow where the cylinder rotation is about to change sign or has just done so.

\begin{figure}[t!]
    \centering
    \includegraphics[scale=0.8]{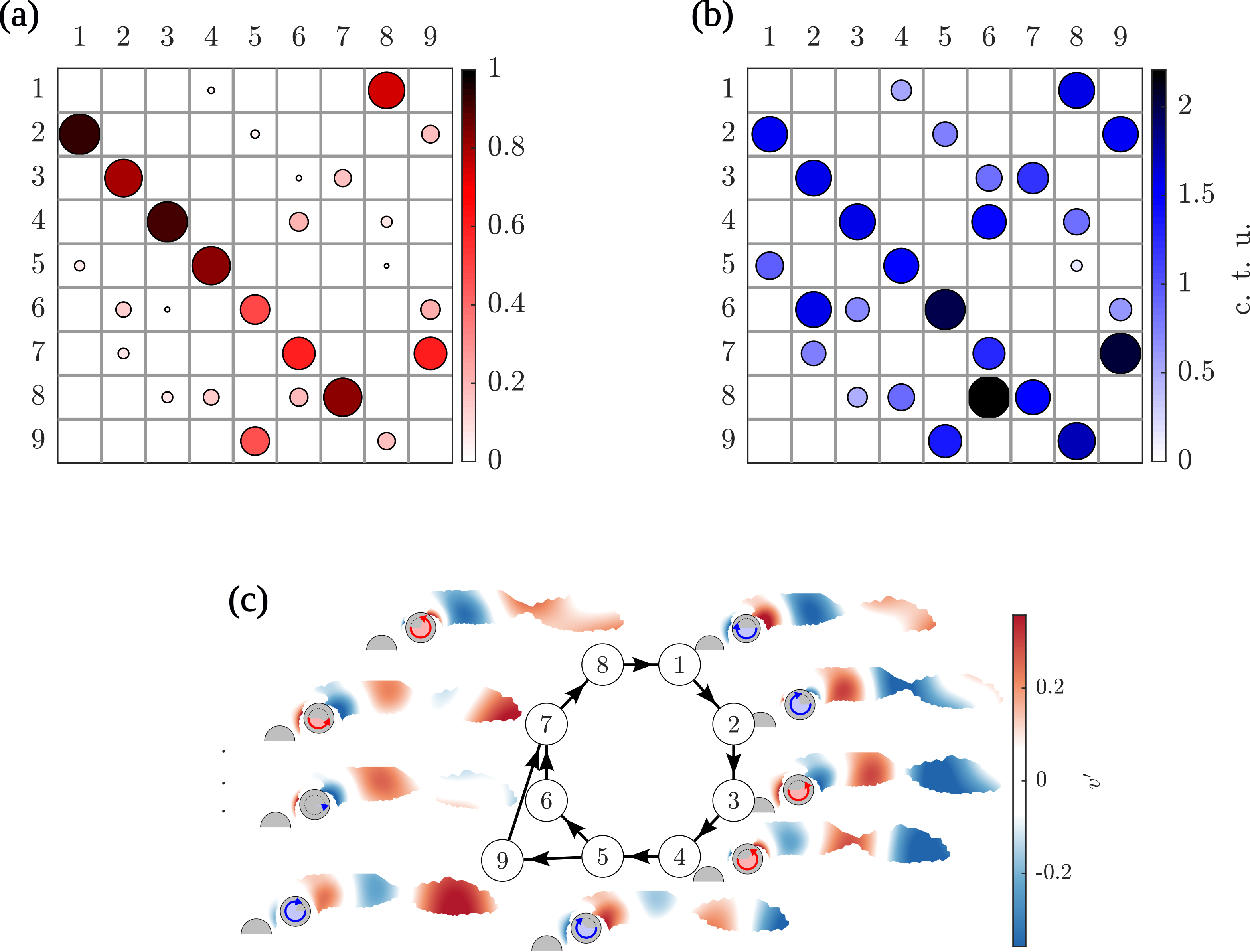}
    \caption{
    Description of subdomain $\mathcal{D}_2$.
    (a) Transition probability matrix;
    (b) Transition time matrix. c. t. u. stands for convective time units.
    (c) shows the main cycle, obtained by following the largest probability transitions.
    }
    \label{Fig:FP_S8}
\end{figure}
Figure~\ref{Fig:FP_S8} details the dynamics of subdomain $\mathcal{D}_2$.
The transition probability unveils two dominant cycles:
$\mathcal{C}_1 \rightarrow
    \mathcal{C}_2 \rightarrow
    \mathcal{C}_3 \rightarrow
    \mathcal{C}_4 \rightarrow
    \mathcal{C}_5 \rightarrow
    \mathcal{C}_6 \rightarrow
    \mathcal{C}_7 \rightarrow
    \mathcal{C}_8$,
and
$\mathcal{C}_1 \rightarrow
    \mathcal{C}_2 \rightarrow
    \mathcal{C}_3 \rightarrow
    \mathcal{C}_4 \rightarrow
    \mathcal{C}_5 \rightarrow
    \mathcal{C}_9 \rightarrow
    \mathcal{C}_7 \rightarrow
    \mathcal{C}_8$.
They display slightly different periods, $12.70$ and $12.84$, respectively, suggesting two different dynamics are at play.
The second one is related to the forcing for two reasons.
First, its period corresponds approximately to $2T_{c1}$.
Second, this cycle includes cluster $\mathcal{C}_9$ with a large actuation level.
The first cycle may result from an interaction between the forcing and the far-field dynamics.
Cluster $\mathcal{C}_5$ is key in switching from one dynamics to the other.

\begin{figure}[t!]
    \centering
    \includegraphics[scale=0.8]{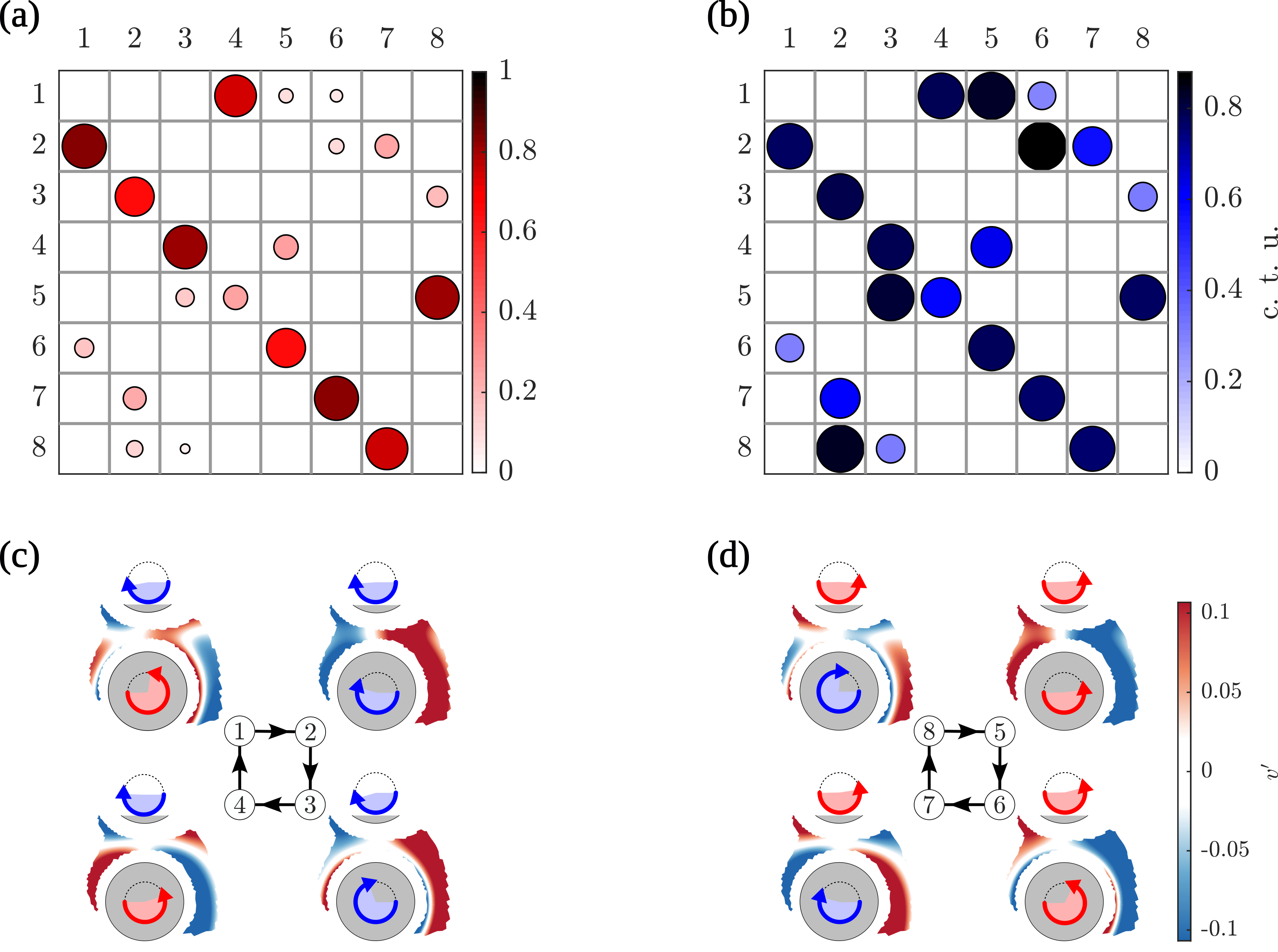}
    \caption{
    Description of subdomain $\mathcal{D}_7$.
    (a) Transition probability matrix;
    (b) Transition time matrix. c. t. u. stands for convective time units.
    The transition probability matrix suggests the dynamics include two cyclic trajectories
    (c) $\mathcal{C}_1 \rightarrow
    \mathcal{C}_2 \rightarrow
    \mathcal{C}_3 \rightarrow
    \mathcal{C}_4$
    and (d)
    $\mathcal{C}_8 \rightarrow
    \mathcal{C}_5 \rightarrow
    \mathcal{C}_6 \rightarrow
    \mathcal{C}_7$.
    The arrows denote the average rotation speed of each cluster, as in Figure~\ref{Fig:FP_S7+S9}.
    }
    \label{Fig:FP_S3}
\end{figure}

Figure~\ref{Fig:FP_S3} details the cluster-based model of subdomain $\mathcal{D}_7$.
Similarly to subdomain $\mathcal{D}_2$, each phase of the control is represented by two clusters,
corresponding to the increase and decrease of the actuation level.
Two cycles can be inferred from the transition probability matrix:
$\mathcal{C}_1 \rightarrow
\mathcal{C}_2 \rightarrow
\mathcal{C}_3 \rightarrow
\mathcal{C}_4$
and
$\mathcal{C}_5 \rightarrow
\mathcal{C}_6 \rightarrow
\mathcal{C}_7 \rightarrow
\mathcal{C}_8$.
The remaining low probability transitions indicate switches from one cycle to another.
A closer inspection of the velocity fields and the rotation speed shows that
the two cycles represent the same behaviour with opposite signs.
Cycles $\mathcal{C}_1$ and $\mathcal{C}_2$ describe the clockwise and counter-clockwise rotation of the top cylinder, respectively.
The four clusters in each cycle describe the rotation of the bottom cylinder.
Two clusters describe the clockwise rotation
($\mathcal{C}_2$ and $\mathcal{C}_3$ for $\mathcal{C}_1$
and
$\mathcal{C}_7$ and $\mathcal{C}_8$ for $\mathcal{C}_2$)
and the other two, the counter-clockwise rotation.
The transition time matrix shows that $\mathcal{C}_1$ and $\mathcal{C}_2$ have different periods, $3.17$ and $3.11$, respectively.
This difference is explained by the incommensurability of $f_{c1}$ and $f_{c2}$,
i.e., $T_1$ and $T_2$ do not have a common measure.




As for subdomains $\mathcal{D}_4$, $\mathcal{D}_5$, $\mathcal{D}_6$, $\mathcal{D}_8$, and $\mathcal{D}_9$, they all describe a periodic behaviour on the surface of the cylinders. They are further detailed in Appendix~\ref{App:FP_TwoCluster}.

The proposed methodology isolates regions of the flow, revealing complex local dynamics that cannot be obtained with a global approach.
Indeed, a global network model of the flow gives a single deterministic cycle with period $12.02$, see Appendix~\ref{App:FP_Global}.


\section{Local interpretation of an experimental jet flow} \label{sec:JetFlow}
In this section, the method is validated with an experimental dataset of planar \ac{PIV} of a water jet flow. The objective is to aid the global modelling of the flow by identifying regions where the dynamics are simple and interpretable.

\subsection{Configuration and data description}

\begin{figure}[t!]
    \centering
    \includegraphics[width = 0.8\linewidth]{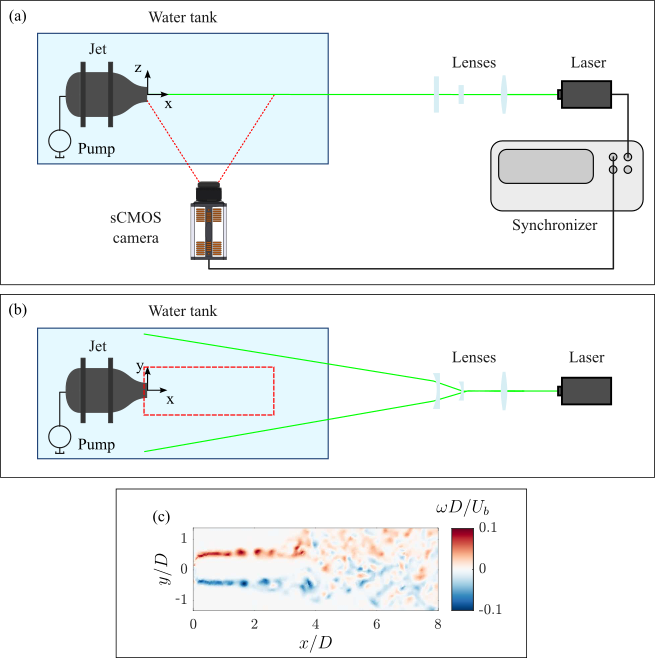}
    \caption{
Schematics of the planar \ac{PIV} setup for the jet flow experiment in water. (a) Top view and (b) front view.
The laser plane, opened with the optical path, is depicted in green,
and the field of view of the \ac{PIV} camera in red.
Figure adapted from \citet{franceschelli2025assessment}, under license CC BY 4.0. (c) Vorticity field of a sample snapshot.
}
    \label{Fig:Jet_configuration}
\end{figure}

Planar \ac{PIV} experiments of a water jet flow (see figure \ref{Fig:Jet_configuration}) are conducted in the water tank facility of the Experimental Aerodynamics and Propulsion lab at Universidad Carlos III de Madrid. The tank has full optical access with dimensions $80 \times 60 \times 40 \;\si{cm}^3$. The jet has a nozzle with an exit diameter of $D = 0.03\;\si{m}$ and is operated at a bulk velocity of $U_b \approx 0.11\;\si{m/s}$. The corresponding Reynolds number based on $U_b$ and $D$ is $Re \approx 3300$, corresponding to a transitional jet flow \citep{bogey2006large,fellouah2009reynolds,mi2013reynolds}. The flow is seeded with polyamide particles of $54\;\si{\micro m}$ diameter and is illuminated at the mid-plane of the nozzle exit with a $5\si{W}$ \textit{LaserTree LT-40W-AA} pulse-width modulated laser. The laser beam is shaped into a thin plane of approximately $ 1\si{mm}$ thickness using a spherical converging lens and two cylindrical diverging lenses, as depicted in figure~\ref{Fig:Jet_configuration}(a). For further insights into the experimental setup, the reader is referred to \citet{franceschelli2025assessment}. Flow snapshots are acquired with an \textit{Andor Zyla sCMOS camera} with a sensor of $5.5$ megapixels ($2160 \times 2560$ px, with a pixel size of $6.5\;\si{\micro m}$). The field of view covers a domain of $8D \times 2.6D$ spanning from the nozzle exit and centred vertically with the centre-line of the jet such that $x/D \in [0,8]$ and $y/D \in [-1.3,1.3]$, nondimensionalised with the jet diameter. The Cartesian coordinate $x$ points in the direction of the jet flow, while $y$ is orthogonal to $x$, pointing upward in the vertical direction. The velocity fields are correspondingly nondimensionalised with the bulk velocity.
The field of view is delimited by a red dashed rectangle in figure~\ref{Fig:Jet_configuration}(b). Finally, figure~\ref{Fig:Jet_configuration}(c) shows the nondimensional vorticity field at a given instant. The small upward deflection is due to asymmetry in the top-bottom boundary conditions.


The dataset consists of a sequence of $60000$ time-resolved snapshots captured with a time step of $\Delta t = 0.011\;\si{s}$.
The \ac{PIV} images are processed with a custom-made multi-frame approach, based on sliding cross-correlation of a 5-frame stencil~\citep{scarano2010advection}.
The processing consists of three-pass cross-correlation analysis with the final interrogation windows of $32\times 32$ px with $75\%$ overlap.
To remove measurement noise, the velocity fields are smoothed with a second-order polynomial Savitzky-Golay filter with a $5\times 5$ kernel in space, and $5$ snapshots in time. 
For the demonstration of the methodology,
$5000$ snapshots, subsampled by a factor of $4$, are considered.
This choice alleviates the computational load by obtaining converged statistics with a reduced number of snapshots.

The jet flow at $\Rey \approx 3300$ corresponds to a transitional jet that is characterised by the presence of localised dynamics in the flow due to the generation and breakdown of ring vortices and turbulent mixing~\citep{fellouah2009reynolds,ball2012flow}.
This compound picture of the flow motivates our choice of the jet flow for the experimental validation of the proposed method.

\subsection{Spatial manifold and clustering results}
\begin{figure}[t!]
    \centering
    \includegraphics[width = \textwidth]{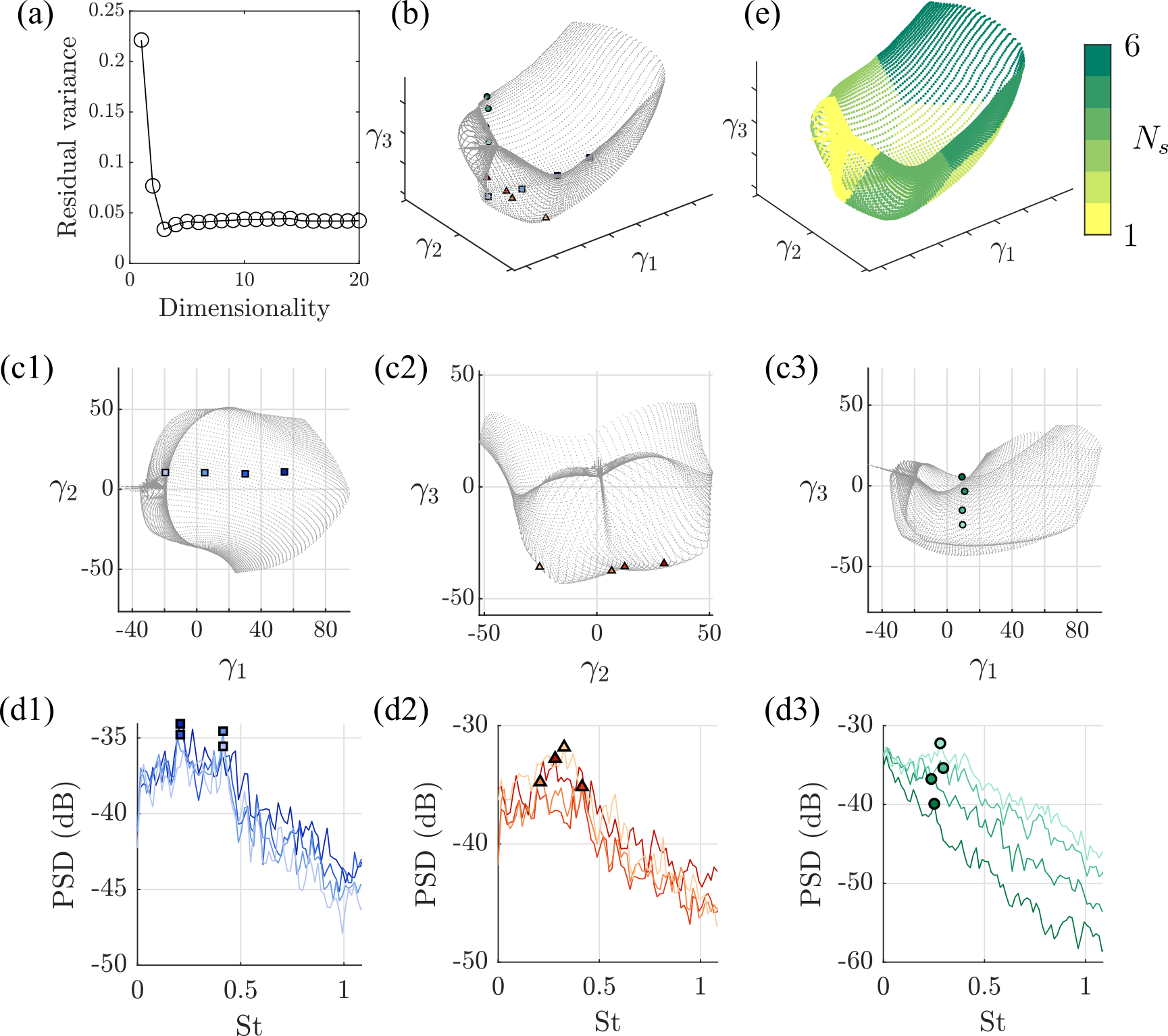}
    \caption{
Spatial manifold of the jet flow data.
(a) Residual variance versus the dimensionality of the spatial reconstruction.
(b) Three-dimensional spatial manifold. Blue coloured squares mark points with $\gamma_2$ and $\gamma_3$ approximately constant ($\gamma_2 \approx 10$ and $\gamma_3 \approx -35$). Orange coloured triangles mark points with $\gamma_1$ and $\gamma_3$ approximately constant ($\gamma_1 \approx -5$ and $\gamma_3 \approx -36$). Green coloured circles mark points with $\gamma_1$ and $\gamma_2$ approximately constant ($\gamma_1 \approx 10$ and $\gamma_2 \approx 49$). (c1-3) 2D views of the spatial manifolds with sampling points. (d1-3) \ac{PSD} of the $v$-velocity at the sampled points. (e) Clustered manifold with $6$ subdomains. 
}
    \label{Fig:JetSpatialManifold}
\end{figure}

To identify regions with similar features and dynamics, the spatial manifold of the vorticity fields is computed with ISOMAP according to the methodology presented in \S\ref{sec:methodology}.
The first three manifold coordinates are selected for the low-dimensional representation as they minimise the residual variance computed according to Equation~\ref{Eq:ResidualVariance}.
The resulting three-dimensional manifold resembles a space coffee cup-like shape and is depicted in figure~\ref{Fig:JetSpatialManifold}. 

To interpret the manifold coordinates, four points are sampled along each direction in the manifold space (see figure \ref{Fig:JetSpatialManifold}(b-d), and the corresponding \ac{PSD} of the $v$ velocity component is computed. 
Interestingly, 
$\gamma_1$ and $\gamma_2$ seem to be correlated with the streamwise and spanwise direction of the flow.
This is partially supported by the distribution of clusters (figures~\ref{Fig:JetSpatialManifold}e and \ref{fig:Jet_overall}), e.g., subdomains $\mathcal{D}_1$ (yellow) and $\mathcal{D}_6$ (darkest green) are located at low and large values of $\gamma_1$, respectively.
The \ac{PSD} of the points distributed along $\gamma_1$ shows clearly two different dominant frequencies, meaning that their dynamics change across this regime. Additionally, it can be noted that the intensity of the \ac{PSD} depends directly on the value of $\gamma_1$, thus increasing for increasing values of $\gamma_1$.
As for $\gamma_2$, the points sampled far from $\gamma_2=0$ seem to include a peak at $St=0.30$ ($f\approx 1.1\;\si{Hz}$) that is mitigated as $\gamma_2$ decreases in absolute value. Additionally, points that are clustered together in the same subdomain retrieve similar dominant frequencies, as can be noticed in figure \ref{Fig:JetSpatialManifold}(d1). On the other hand, $\gamma_3$ correlates well with the \ac{PSD} intensity. Namely, moving towards a more negative $\gamma_3$ results in a higher intensity of the \ac{PSD}. Lastly, in figure \ref{Fig:JetSpatialManifold}(e), we note a symmetry along the $\gamma_2$ direction equivalent to the symmetry with respect to the $x$ axis. Close to the symmetry plane, the dominant frequencies are similar, while moving away, the dominant frequencies show similar intensities.
The symmetry of the configuration explains the symmetry of the spatial manifold along the $\gamma_2$ axis.
The deviation from perfect symmetry may be explained by the deflection of the jet towards the free surface of the water tank.


The spatial manifold is clustered into 6 subdomains following the \ac{TLF} criterion.
The corresponding clustered domain is depicted in figure~\ref{fig:Jet_overall}, where the subdomains are ordered according to their total kinetic energy from larger to smaller. It is possible to distinguish two main regions. The first is the initial region, which spans the length of the potential core (subdomain $\mathcal{D}_1$). The second region consists of several clusters (subdomains $\mathcal{D}_2$, $\mathcal{D}_3$, $\mathcal{D}_4$, $\mathcal{D}_5$, $\mathcal{D}_6$) and corresponds to the transitional region. 

In the literature, these regions have been identified for submerged free round jet flows~\citep{Abramovich1963,yule1978large}. The initial region covers the potential core, within which the jet centreline remains approximately equal to the exit velocity due to limited penetration of the surrounding shear layer. This region typically extends over $4$–$6D$, while the shear layer grows inward from the nozzle lip towards the jet centreline. 

The potential core is surrounded by a mixing layer that develops from the nozzle edges due to the strong velocity gradient between the jet exit and the ambient fluid ($U_{\infty} \approx 0$ in a quiescent environment). These velocity gradients give rise to Kelvin–Helmholtz instabilities, which roll up into vortical structures (vortex rings in axisymmetric jets). These structures enhance turbulent mixing and entrain ambient fluid into the jet, leading to progressive growth of the shear layer and the gradual reduction of the potential core along the axial direction until it disappears (see Appendix \ref{App:JetSubdomains}).

It is important to note that the core, the shear layer, and ambient flow are strongly coupled through turbulent entrainment and momentum exchange. The shear layer entrains ambient fluid toward the jet core, which contributes to a reduction of centreline momentum and velocity, thereby controlling the rate at which the potential core decays. At the same time, the velocity difference across the shear layer provides the energy that sustains Kelvin–Helmholtz instabilities and the growth of turbulent structures. The shear layer thus acts as the main interface between the jet core and the surrounding fluid, where coherent vortical structures mediate the exchange of momentum and mass between the inner and outer regions. Because of this strong coupling, this region is identified as a single cluster in the spatial partitioning. However, when increasing the number of clusters, the shear-layer region is typically the first to be further subdivided, reflecting its strong internal gradients and dynamical complexity.

Downstream of the initial region, the transitional region begins, in which the jet continues to develop and typically extends up to $5$–$10D$. In this region, the velocity profile evolves significantly without reaching full self-similarity; it therefore acts as an adjustment zone before the jet enters the fully developed region. Consequently, different clusters along the streamwise direction can be identified, each representing successive stages of mixing and momentum redistribution in the developing jet. When increasing the number of clusters, this region presents a larger partition into smaller clusters, describing this mixing and evolving phenomenon.

Note, that this framework enables the identification of kinematically coherent regions even when they are spatially separated, meaning that the partitioned subdomains do not necessarily need to be spatially together. In Appendix~\ref{App:JetSubdomains}, the effect of a larger and smaller number of subdomains, and the spatial separation between subdomains, is assessed.
The comparison between the clustered manifold and the clustered domain
suggests that $\gamma_1$ and $\gamma_2$ are correlated with the physical coordinates of space.
This represents an interesting feature of the method, since it is capable of identifying a spatial distribution with clear interpretability in the physical space, without introducing explicitly any information about the real coordinates. 

Furthermore, it is interesting to note that, similar to the pinball, the largest domain (subdomain $\mathcal{D}_6$) is condensed into a small, clustered area in the manifold. A reason behind this might be the clear dominance of a dynamics, which in this region corresponds mainly to the shear layer where the generated vortices roll up and then pair up \citep{ball2012flow}. On the other hand, in regions where complexity increases due to mixing and turbulence, the corresponding manifold areas are seen to be larger as the dynamics are spectrally broadband.

\subsection{Local cluster-based modeling}


\begin{figure}[htbp!]
    \centering
    \includegraphics[width = \textwidth]{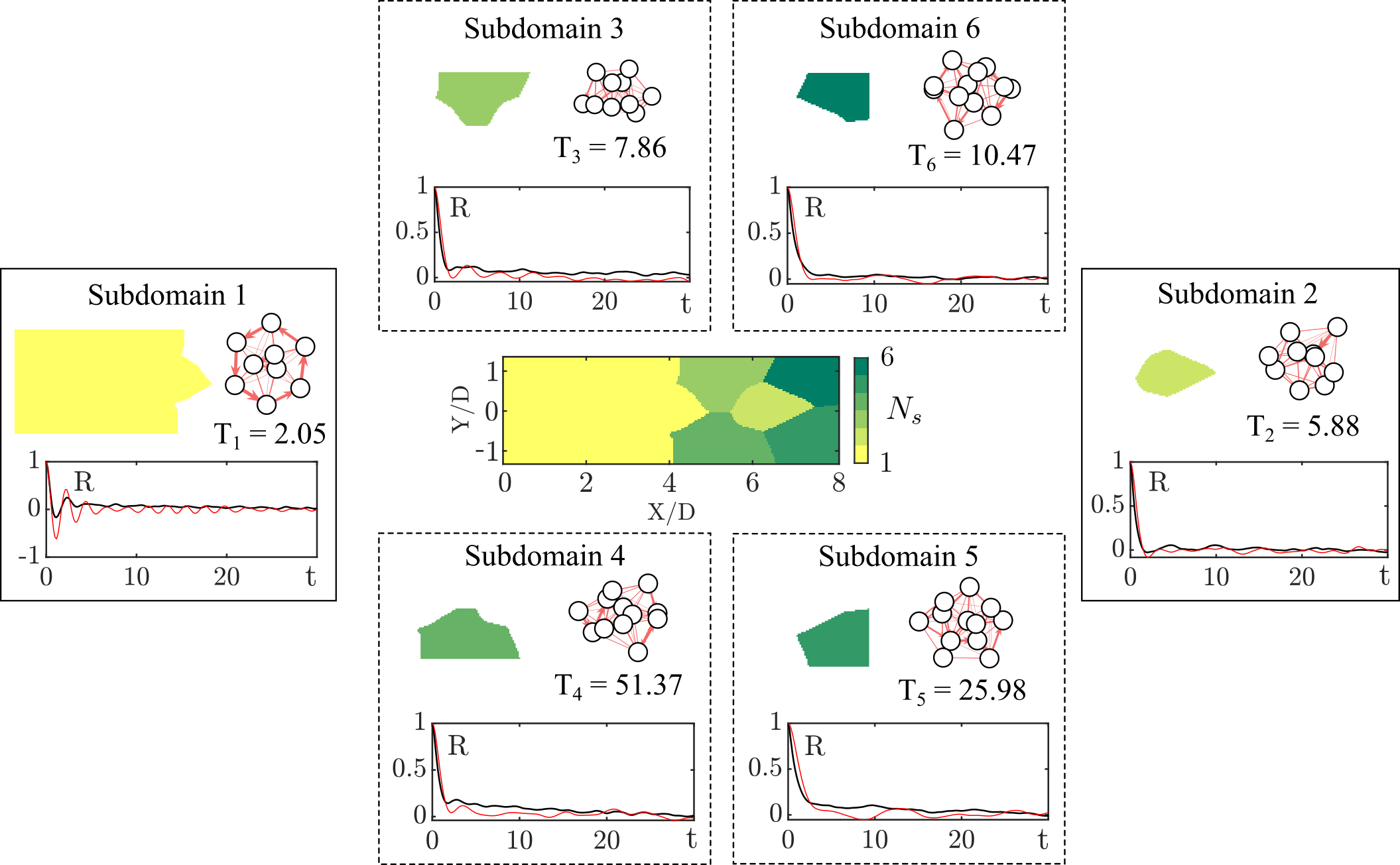}
    \caption{Local interpretation of the experimental water jet flow. The flow is partitioned into $6$ subdomains. Each subdomain is accompanied by the cluster-based network model built on the masked snapshots and the autocorrelation functions of the original snapshots and the reconstructed snapshots as a function of the time nondimensionalised with the convective time. The dots correspond to the centroids in the clusters projected on the first two \ac{POD} modes. The times $T_i$ correspond to the first zero of the autocorrelation function of the reconstructions. The autocorrelation functions are depicted for the predicted data (red line) and the original data (black line).
    }
    \label{fig:Jet_overall}
\end{figure}

The analysis of the dynamics in each of the spatial domains is done through cluster-based modelling. Figure~\ref{fig:Jet_overall} shows an exploded view of the jet flow into the six subdomains. 

Note that the flow exhibits asymmetry due to boundary effects in the water tank, since the experimental jet is not fully canonical and is affected by recirculation and pressure-gradient imbalances, leading to a deflected jet in the vertical direction. 
The clustering method is able to capture this asymmetry at the level of the clustering, resulting in a partition that reflects the global imbalance of the flow.

Then, for each subdomain, the cluster-based models are extracted, with the corresponding number of temporal clusters selected according to the two-line fit criteria.
For each of these cases, the snapshots are reconstructed by interpolating between the cluster centroids.
The autocorrelation between the reconstructed snapshots and the original ones is used to assess whether the local models are representative of the dynamics. Results show that only subdomains $\mathcal{D}_1$ and $\mathcal{D}_2$ present similar periods (first zero crossing point), thus the analysis and discussion of the results are only assessed for these two subdomains. All periods are summarised in table~\ref{Tab:Jet_Periods}.

\begin{table}
    \centering
   \begin{tabular}{lcccccc}
    \toprule
        Cluster \# & 1 & 2 & 3 & 4 & 5 & 6\\
        \midrule
        Original data & 2.67 & 5.91 & 130.48 & 130.48 & 115.03 & 72.14\\
        \midrule
        CNM prediction & 2.05 & 5.88 & 7.86 & 51.37 & 25.98 & 10.47\\
        \bottomrule
    \end{tabular}
    \caption{Dominant nondimensional period for each subdomain of the jet flow obtained via the first zero of the autocorrelation function.}
    \label{Tab:Jet_Periods}
\end{table}

In the other subdomains, high turbulent mixing with three-dimensional effects is observed due to chaotic behaviour and turbulent dynamics, rendering dynamic modelling impractical.

\begin{figure}[htbp!]
    \centering
    \includegraphics[width = 0.8\textwidth]{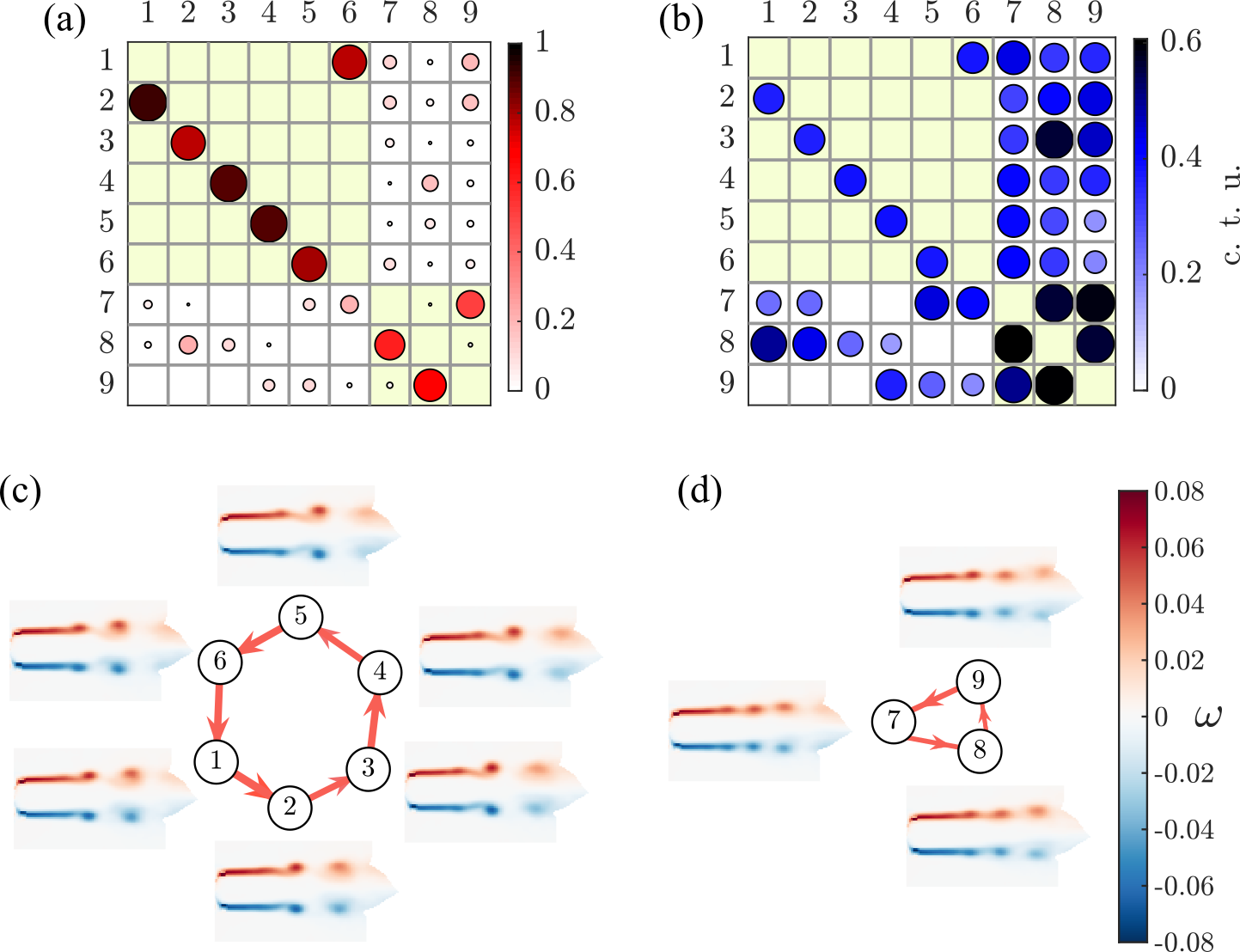}
    \caption{Description of subdomain $\mathcal{D}_1$. a) Transition probability matrix. b) Transition time matrix. The transition probability matrix suggests that dynamics include two periodic trajectories ($\mathcal{C}_1 \rightarrow \mathcal{C}_2 \rightarrow \mathcal{C}_4 \rightarrow \mathcal{C}_5 \rightarrow
    \mathcal{C}_6$ and $ \mathcal{C}_7 \rightarrow \mathcal{C}_8 \rightarrow \mathcal{C}_9 $). The corresponding nondimensional vorticity snapshots of the centroids and the subnetworks are displayed in c) and d). Limit cycle 1 has period $T_{1,1} = 2.29$ and cycle 2 of period $T_{1,2} = 1.84$.}
    \label{Fig:Jet_domain6}
\end{figure}
Finally, figure~\ref{Fig:Jet_domain6} presents the results of the local Markov models, showing the transition probability matrix with the corresponding transition time matrix. For subdomain $\mathcal{D}_1$, the cluster model indicates a switch between two phases, that can be identified from the transition probability matrix ($\mathcal{C}_1 \rightarrow \mathcal{C}_2 \rightarrow \mathcal{C}_3 \rightarrow \mathcal{C}_4 \rightarrow \mathcal{C}_5 \rightarrow \mathcal{C}_6$ and $ \mathcal{C}_7 \rightarrow \mathcal{C}_8 \rightarrow \mathcal{C}_9 $) with periods $T_{1,1} = 2.29$ and $T_{1,2} = 1.84$, respectively.
The first phase is characterised by two large vortices,
while the second displays three small vortices.
Both cycles describe the dynamics of the vortex shedding mechanism, with the difference that the first one captures the dynamics post-vortex pairing, similarly to the work of \citet{Kaiser2014jfm}, where the authors identify two cycles in a mixing layer: one related to the Kelvin-Helmholtz instability and the other to a vortex pairing.

These dynamics are confirmed with a refined model based on 20 clusters, see Appendix~\ref{App:JetSubdomains}.
This high-resolution model identifies a centroid where two vortices merge that serves as a transition between the two cycles. However, this analysis is presented for the sake of interpretation and not for the identification of limit cycles. Increasing the number of clusters highly reduces the number of snapshots per cluster, thus increasing the variability of results despite augmenting the number of replicates in the $k$-means algorithm.
The transition between both phases is between $\mathcal{C}_2$ and $\mathcal{C}_7$ in the low-resolution model. 

Moreover, the difference in frequency might be induced by instabilities of the flow coming from recirculation, asymmetry due to the open tank, non-zero pressure gradient, or three-dimensional effects, which cause the vortex shedding to trigger earlier. However, such a difference only comprises a $20\%$ change in the period and corresponding wavelength, which is compared between the vorticity fields of the centroids of $\mathcal{C}_5$ and $\mathcal{C}_8$. These are selected to be the instant when the first vortex is fully detached. The bulk velocity for both cycles remains unchanged. Evaluating the probability of being in each of the cycles, it is observed that there is the same probability for both cycles, where a regular transition between cycles is observed in a quasi-periodic manner with a period equal to $T_{1}$. 

\begin{figure}[htbp!]
    \centering
    \includegraphics[width = 0.8\textwidth]{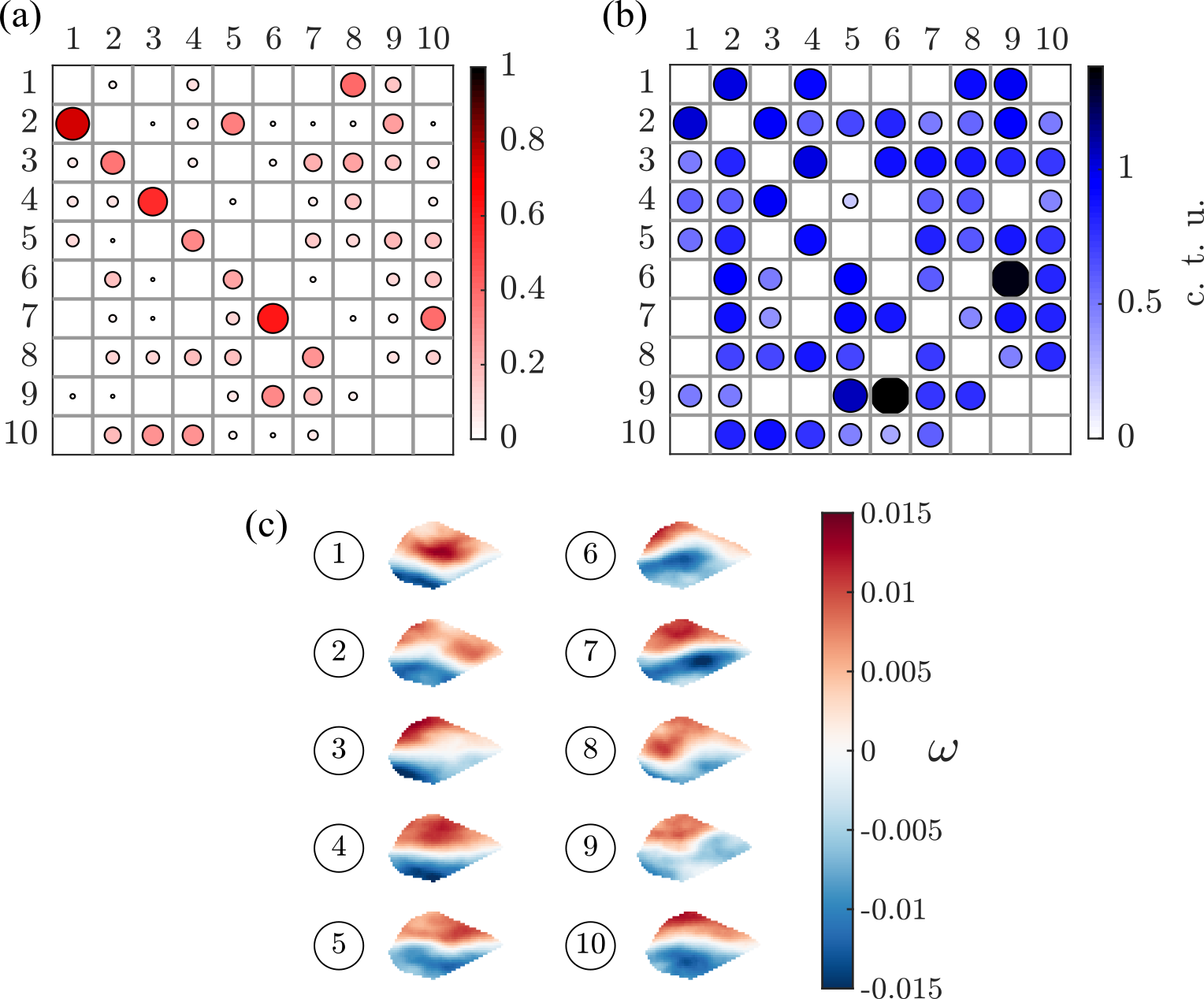}
    \caption{Description of subdomain $\mathcal{D}_2$. (a) Transition probability matrix. (b) Transition time matrix c. t. u. stands for convective time units. Transition probability matrix suggests the dynamics include a periodic trajectory ($\mathcal{C}_2 \rightarrow \mathcal{C}_3 \rightarrow \mathcal{C}_4 \rightarrow \mathcal{C}_5 \rightarrow
    \mathcal{C}_6 \rightarrow
    \mathcal{C}_7 $). The period of the cycle is $T = 6.07$. (c) Nondimensional vorticity snapshots of the cluster centroids.
}
    \label{Fig:Jet_domain5}
\end{figure}
The results of subdomain $\mathcal{D}_2$ are represented in figure \ref{Fig:Jet_domain5}, where a total of $10$ clusters have been identified. Different from before, no dominant cycles are identified, but instead, many interconnected cycles can be observed. In this subdomain, flow oscillates in a switching pattern, as can be seen in the vorticity fields of the corresponding centroids (see figure~\ref{Fig:Jet_domain5}(c). This waving behaviour is generated by the collision of vortices in the centreline and the three-dimensional interaction of the flow. The result is an oscillating pattern that repeats in time but can follow different paths, i.e., different cycles. It is important to note that due to the high variability of the cycles and the transitions, the $k$-means clustering was performed with an increased number of replicates, of the order of a thousand, in order to ensure convergence in the cluster transitions.

In summary, the present methodology identifies local dynamics that are not identified with a global approach. The global cluster-based analysis does not properly identify the separate dynamics observed in subdomain $\mathcal{D}_1$ nor the oscillation of subdomain $\mathcal{D}_2$. The dynamics are mixed together without retrieving a clear separation between them, see Appendix~\ref{App:Jet_Global}.


\section{Discussion and conclusions}\label{Sec:Conclusion}

This paper introduces \ac{ST-CNM}, a method to identify local dynamics that are not captured by global methods.
The starting point is a space and time-resolved snapshot dataset.
The method first consists of the decomposition of the flow domain with an unsupervised method applied directly to the data points.
The key enabler is the definition of a low-dimensional representation of the flow domain that separates the regions based on the local dynamics.
This low-dimensional space, termed spatial manifold, is obtained via a nonlinear decomposition of the vorticity field with \ac{ISOMAP}.
Then, each of the corresponding subdomains is clustered in time, according to the cluster-based network model methodology, to build local Markov models.
The number of spatial and temporal clusters is decided by the \ac{TLF} method, making the process fully automated without the need for any meta-parameter tuning.
The method is demonstrated on two flows: a numerical simulation of the fluidic pinball under control and the PIV data of an experimental jet flow at $\Rey \approx 3300$.

For the fluidic pinball, the flow is forced by a periodic rotation of the rear cylinders at incommensurable frequencies.
ISOMAP identifies a four-dimensional space that represents the spatial distribution of the dynamics.
Interestingly, the first three spatial manifold coordinates are related to the main local frequency, the intensity of the vortex shedding frequency, and the average local vorticity.
The fourth manifold coordinate represents a nonlinear feature.
The $k$-means algorithm applied to the manifold coordinates separates the domain into $9$ subdomains
and identifies the global vortex shedding and local periodic behaviours related to the forcing that are not captured with a global approach.
Intriguingly, the periodic oscillation of the top cylinder is also perceived upstream, on the surface of the front cylinder.
Although the small regions near the cylinders share the same periodic behaviour, they are separated due to their average vorticity level.
Finally, the method identifies two regions with complex dynamics: (1) the region between the two rear cylinders is under the influence of two incommensurable rotations, and (2) the wake of the top cylinder that experiences a nonlinear interaction with the global vortex shedding.
The method suggests that those regions require further modelling effort to accurately represent the local dynamics.

Results of an experimental dataset have been presented for the planar \ac{PIV} of a transitional jet flow. In this case, \ac{ISOMAP} identifies a three-dimensional space to represent the dynamics. In this low-dimensional space, the first and second manifold coordinates are closely related to the spatial coordinates, meaning that the model is able to automatically identify a relation with the real geometry of the flow field without any prior information. On the other hand, the third manifold dimension does not present such a strong relationship but shows a change in the power level of the \ac{PSD}. The partitioning of the domain with $k$-means retrieves $6$ subdomains, maintaining the symmetry of the flow. Firstly, a main subdomain is identified where the shear layer, vortex formation, and pairing occur. The corresponding local Markov model is able to identify and separate the dynamics of the vortex shedding phenomena and the vortex pairing. Downstream of the potential core, a second subdomain is identified where the oscillating dynamics of the flow are identified, and finally, the remaining subdomains are observed to be turbulent with strong mixing. The method is shown to be capable of identifying local dynamics that, otherwise, in a global analysis cannot be directly identified. 

Although we propose a method to select the number of clusters,
the framework allows a description of the flow at virtually all levels of granularity in time and space.
However, the relevance of the description depends on the data resolution.
The proposed method constitutes a diagnostic tool to locate regions with complex dynamics that may require additional modelling efforts as opposed to classical projection-based methods or system identification \citep{brunton2016discovering,brunton2016sparse,peherstorfer2016data,kramer2024learning},
or those that require monitoring with increased time and space resolution.

The automatization of the process is key to include the method in optimisation loops such as design and control, to reduce recirculation regions on complex geometries, or to mitigate local fluctuations at frequencies that may induce resonance, to give a few examples.
Future work shall focus on deriving the interaction between the subdomains to build a cluster-based global model of the flow that takes into account the local dynamics and the interaction between subdomains.
Another potential line of development is the description of nonstationary flows and the time evolution of the subdomains.


\section{Acknowledgements}
This project has received funding from the European Research Council (ERC) under the European Union’s Horizon 2020 research and innovation program (grant agreement No 949085). Views and opinions expressed are, however, those of the authors only and do not necessarily reflect those of the European Union or the European Research Council. Neither the European Union nor the granting authority can be held responsible for them.
We appreciate fruitful discussions with L. Franceschelli, L. Marra, M. Raiola, and I. Tirelli.

\appendix
\setcounter{figure}{0}
\setcounter{table}{0}

\section{Interpretation of the two-cluster models}\label{App:FP_TwoCluster}
\counterwithin{figure}{section}
\counterwithin{table}{section}

\begin{figure}[htbp!]
    \centering
    \includegraphics[width = \textwidth]{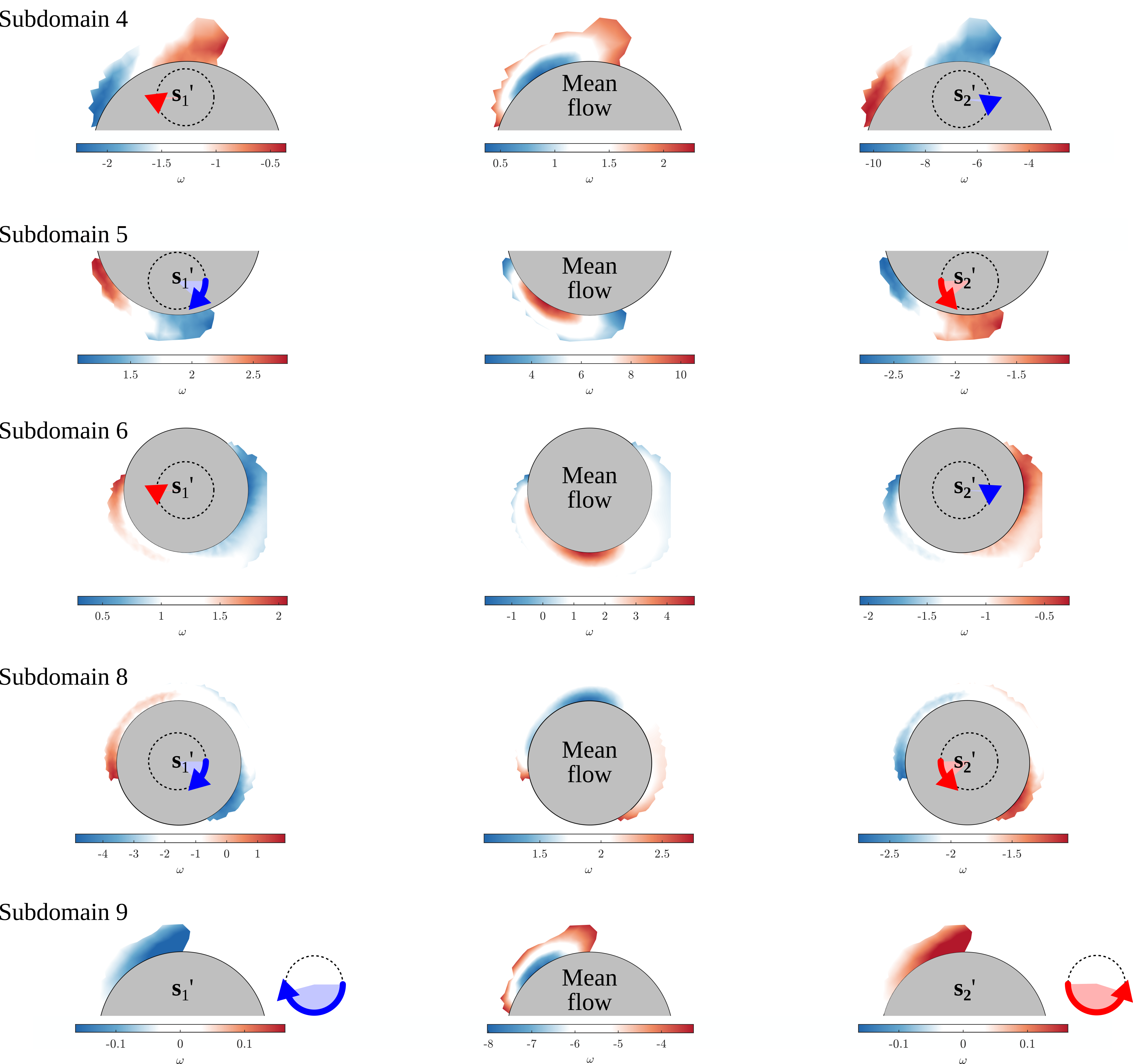}
    \caption{Description of the network model with only two clusters.
    For each subdomain, the vorticity of the mean flow is displayed in the central column and the fluctuations in the left and right columns.
    The first subdomain is located on the surface of the front cylinder that does not rotate.
    It is under the influence of the top cylinder.
    }
    \label{Fig:FP_SDuo}
\end{figure}
Figure~\ref{Fig:FP_SDuo} shows the snapshots for the subdomains with only two clusters.
They all demonstrate a periodic behaviour
with a pocket of intense vorticity on the surface of the cylinder that moves following the rotation direction.
The pocket of vorticity is rather stable for subdomain 9, as it lies on the front cylinder that does not rotate.
However, it is still affected by the rotation of the top cylinder.
A symmetric pocket of vorticity on the front cylinder is identified for a larger number of subdomains.

\section{Global CNM of the fluidic pinball under control}\label{App:FP_Global}
\counterwithin{figure}{section}
\counterwithin{table}{section}

\begin{figure}[htbp!]
    \centering
    \includegraphics[width = \textwidth]{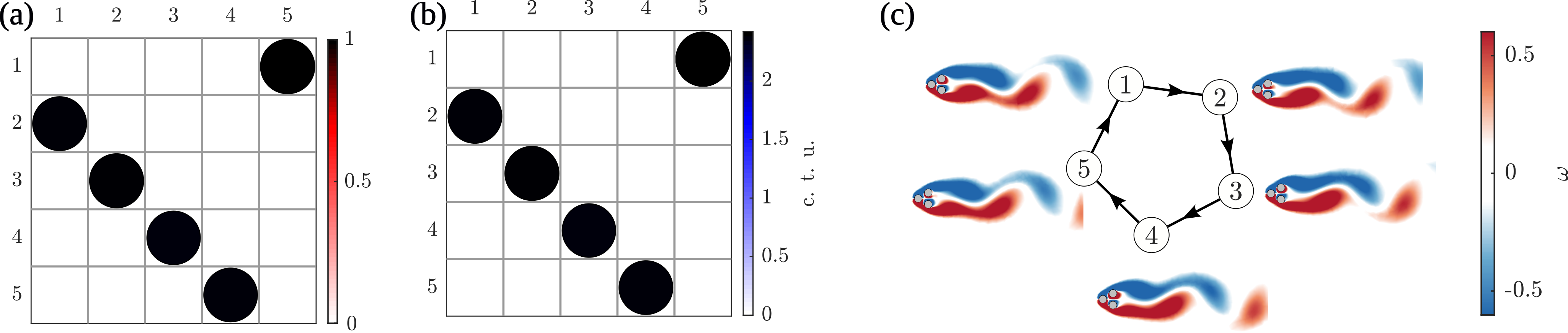}
    \caption{(a) Transition probability matrix,
    (b) Transition time matrix and cluster-based model for the fluidic pinball under control as a whole.
    }
    \label{Fig:FP_Global}
\end{figure}

Figure~\ref{Fig:FP_Global} shows the global CNM of the fluidic pinball under periodic control as described in \S~\ref{Sec:FluidicPinball}.
The optimal number of clusters following the \ac{TLF} method is 5.
The transition probability network reveals a deterministic transition from one cluster to the other.
The transition time from one cluster to the other is similar for all clusters and is approximately 2.4 convective time units.
The sequence of vorticity snapshots shows that CNM identifies the dynamics of the far-field vortices, i.e., the shedding of large vortices due to the pairing of the vortices shed by the top and bottom cylinders.


\section{Local Markov models for jet flow subdomains}\label{App:Jet}
\counterwithin{figure}{section}
\counterwithin{table}{section}

\subsection{Effect of number of subdomains in spatial clustering of jet flow} \label{App:JetSubdomains}

\begin{figure}[htbp!]
    \centering
    \includegraphics[width = \textwidth]{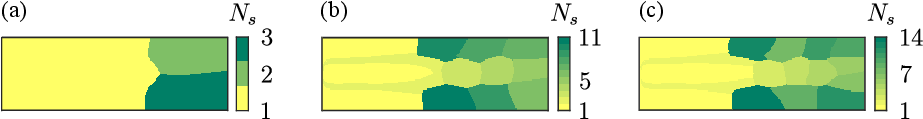}
    \caption{Spatial partitioning of jet flow with different numbers of subdomains: 3, 11, and 14.}
    \label{fig:Jet_SpatialClusters}
\end{figure}

In this section, the effect of the number of subdomains selected for spatial clustering on the spatial distribution of the flow field is assessed. Four different cases have been considered: 3, 11, and 14 subdomains. The corresponding clustered domains are displayed in Figure \ref{fig:Jet_SpatialClusters}. 

It can be noted that the flow starts clustering into subdomains in regimes where the complexity of the flow highly increases. In the case of the jet flow, this is found downstream of the jet, where high turbulence intensity and mixing take place. Then, when the number of clusters is increased, the flow is clustered further upstream until reaching the optimal clustering with 6 subdomains obtained from the \ac{TLF} method. Further increasing the number of clusters results in a higher partitioning far downstream in the flow field, but no new clusters are identified. Note that with 11 clusters, the results show that it is possible to isolate the dynamics of the shear layer into a unique cluster, and this structure persists as the number of clusters increases. Moreover, it is interesting to observe that in the case of $14$ clusters, the subdomain $\mathcal{D}_2$ appears as two disconnected regions. This behaviour is expected since, in the full three-dimensional flow, this region is connected. However, the planar representation of the flow fields, without out-of-plane information, leads to this partitioning into separated subdomains. 
More generally, in different flow configurations, regions that are not spatially connected but exhibit similar dynamics may also be grouped together. This is consistent with the objective of the method, which is to partition the flow into subregions whose dynamics can be effectively represented by interpretable reduced-order models.

\subsection{Analysis of dynamics of jet flow subdomain $\mathcal{D}_1$}
\begin{figure}[htbp!]
    \centering
    \includegraphics[width = 0.9\textwidth]{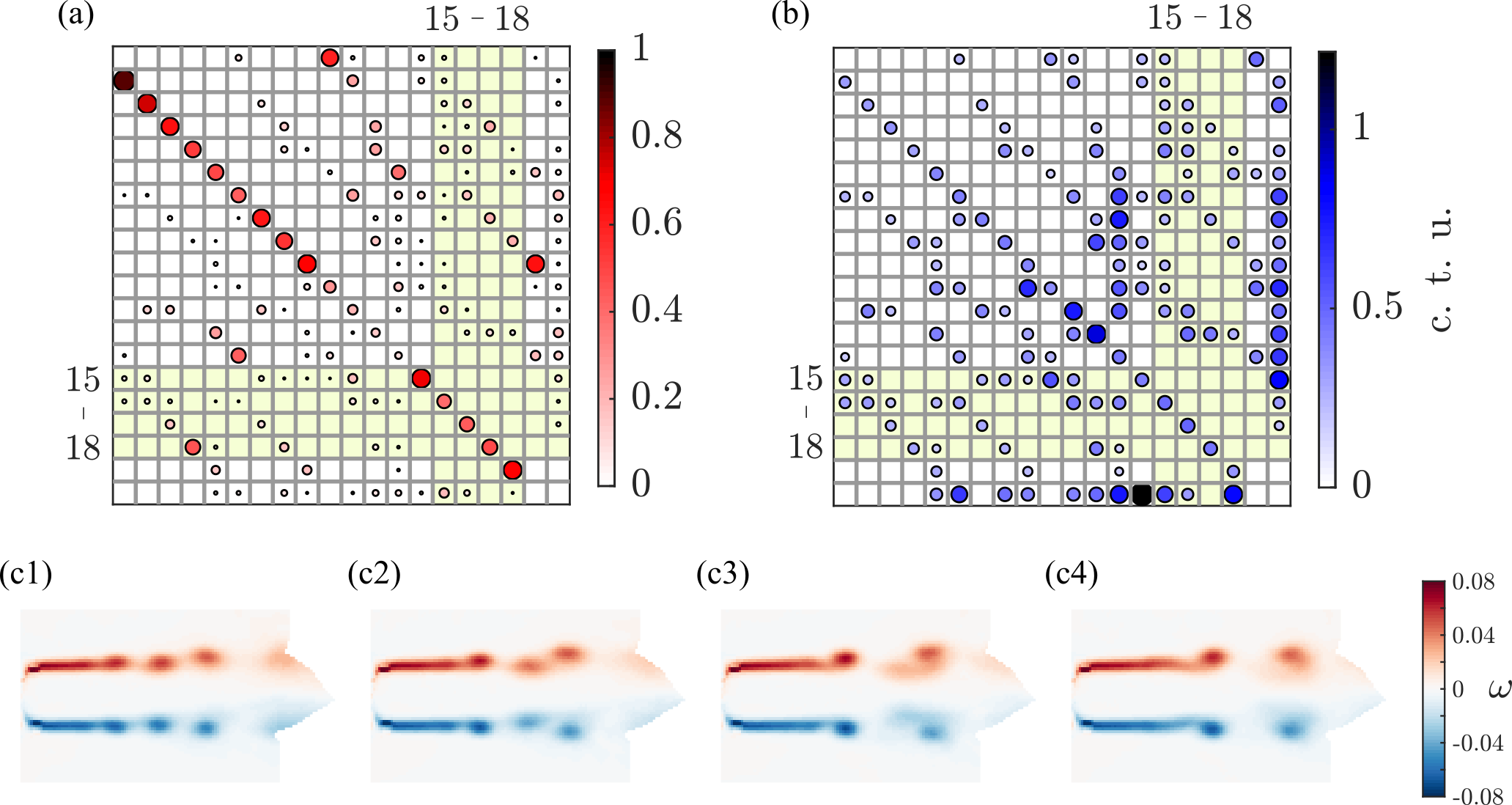}
    \caption{Description of subdomain $\mathcal{D}_1$ with 20 temporal clusters. a) Transition probability matrix. b) Transition time matrix c. t. u. stands for convective time units. c) Nondimensional vorticity snapshots at the centroids of $\mathcal{S}_{15}-\mathcal{S}_{18}$ corresponding to the transition between the vortex shedding and the vortex pairing.}
    \label{fig:Jet_subdomain6_C20}
\end{figure}

This appendix presents the analysis of the dynamics in subdomain $\mathcal{D}_1$ with a larger number of temporal clusters. To increase the resolution of the dynamics extracted from the local analysis, 20 clusters are used to compute the temporal clustering. Results are presented in figure \ref{fig:Jet_subdomain6_C20}. Note that, unlike before, the interpretability of the results is now more complex, as no clear cycles are identified in the transition probability maps. Additionally, when analysing the transition between centroids, it is observed that the two cycles previously obtained with 9 clusters are now intermixed, indicating a lack of a smooth transition between centroids. This loss of smoothness in the reconstruction of dynamics is due to the small number of samples per cluster, which is not enough to resolve the dynamics within a large number of clusters. However, this extended analysis with more clusters allows identification of the transition snapshots between vortex shedding and the vortex pairing effect. This mechanism is clearly depicted in figure \ref{fig:Jet_subdomain6_C20}c) where the vorticity field at the centroids of the transition is represented and a leapfrog-like mechanism is observed.

\section{Global CNM of the jet flow}\label{App:Jet_Global}

\begin{figure}[htbp!]
    \centering
    \includegraphics[width = 0.9\textwidth]{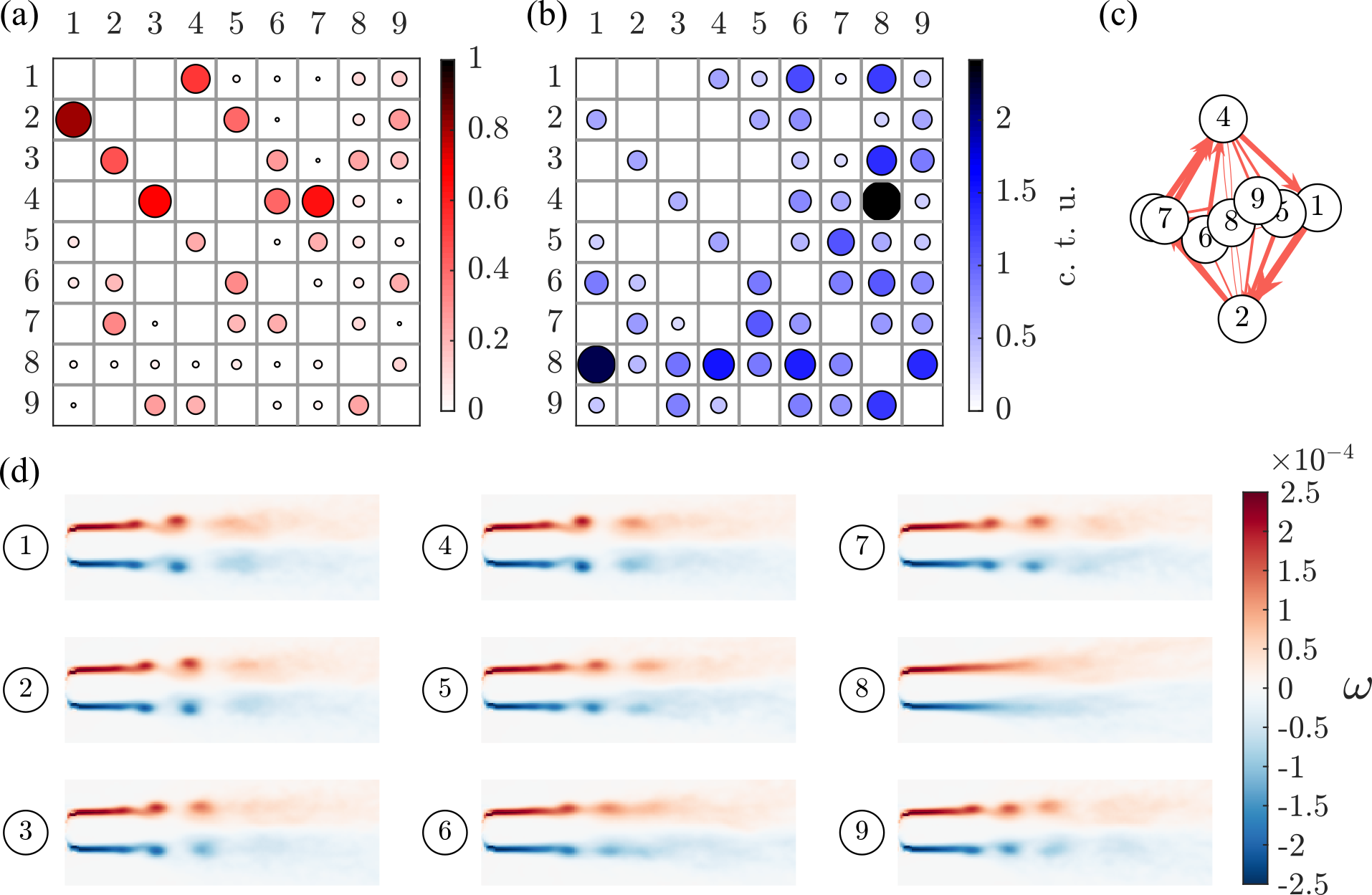}
    \caption{Jet flow global CNM. (a) Transition probability matrix. (b) Transition time matrix. (c) Cluster centroid map network. (d) Nondimensional vorticity snapshots of the centroids of the clusters.
    }
    \label{Fig:JetGlobal_network}
\end{figure}

In this section, the global \ac{ST-CNM} of the jet flow is discussed as introduced in Section~\ref{sec:JetFlow}. Figure~\ref{Fig:JetGlobal_network} shows the results of the global analysis of the jet with 9 clusters. A periodic trajectory can be identified between $\mathcal{C}_1 \rightarrow \mathcal{C}_2 \rightarrow \mathcal{C}_3 \rightarrow \mathcal{C}_4$ representing the vortex shedding mechanism that was observed in the local analysis of subdomain $\mathcal{D}_1$. The other centroids do not show a clear trajectory as observed in figure~\ref{Fig:Jet_domain6}, but they represent a transition within the vortex shedding phenomena. Additionally, it is remarkable to mention that no clear dynamics are identified in the region corresponding to subdomain $\mathcal{D}_2$. In the global analysis, the dominant dynamics that can be analysed are those found in subdomain $\mathcal{D}_1$, where only the vortex shedding can be identified. Thus, all the remaining information is grouped together as transition points, as can be seen in the network of figure~\ref{Fig:JetGlobal_network} c).

\bibliographystyle{unsrtnat}
\bibliography{references}  






\end{document}